\title{Investigating Wheat Price with a Multi-Agent Model}

\documentclass[11pt]{amsart}

\pdfoutput=1		

\usepackage{graphicx}
\usepackage{amssymb}
\usepackage{geometry}                
\title{Investigating Wheat International Market with a Multi-Agent Model}

\author[G. Giulioni]{Gianfranco Giulioni}
\thanks{Department of Philosophical, Pedagogical and Economic-Quantitative Sciences, ``G. D'Annunzio'' University, Pescara (Italy)}
\author[E. Di Giuseppe]{Edmondo Di Giuseppe}
\author[M. Pasqui]{Massimiliano Pasqui}
\author[P. Toscano]{Piero Toscano}
\author[F. Miglietta]{Francesco Miglietta}
\thanks{National Research Council-Institute of Biometeorology (CNR-Ibimet), Roma, Italy}

\email{gianfranco.giulioni@unich.it}

\usepackage{natbib}
\usepackage{hyperref}
\usepackage{diagbox}
\usepackage{array}
\newcolumntype{M}[1]{>{\centering\arraybackslash}m{#1}}
\usepackage{float}
\usepackage{amssymb,amsmath}
\usepackage{multirow}	

\begin{document}
\maketitle 


\begin{abstract}
In this paper we build a computational model for the analysis of international wheat spot price formation, its dynamics and the dynamics of internationally exchanged quantities.
The model has been calibrated using FAOSTAT data to evaluate its in-sample predictive power. 
The model is able to generate wheat prices in twelve international markets and wheat used quantities in twenty-four world regions. The time span considered goes from 1992 to 2013. In our study, a particular attention was paid to the impact of Russian Federation's 2010 grain export ban on wheat price and internationally traded quantities. Among other results, we find that wheat average weighted world price in 2013 would have been 3.55\% lower than the observed one if the Russian Federation would have not imposed the export ban in 2010.           
\end{abstract}



\section{Introduction}

The worldwide supply of food in addition to the conditions of access to it by individuals are strictly connected to the concept of food security. The Food and Agriculture Organization of United Nations (FAO) identifies the four pillars of food security as availability, access, utilization, and stability. 

In this framework, the volatility of commodity prices of the agricultural market observed in recent years is an issue, so much so that the European and international agricultural policy has shown a clear interest in reducing it effectively. Between 2008 and 2011 there has been changes in price as much as 100\% in a year and the other. The maize (corn) was worth euro 129 per ton (129/t) in July 2006, then euro 283/t in March 2008 (+ 119\%) and euro 139/t in September 2009 (-51\%). In 2011 it rose to euro 290/t (+ 109\%). The price of other cereals has had a similar trend and dramatic swings.

These oscillations are due to several reasons, often complex and sometimes linked to real speculation based on emotion and ignorance of Securities Dealers.

The first and perhaps most important structural element of the agricultural market volatility lies in the inherent fluctuation that is the basis of agricultural production. A good year and a bad year from a climate point of view can have decisive impacts on production levels of a company and/or region, with the markets being quite sensitive to weather information that affects the yield potential of the growing crop. It is important for market operators to be able to predict the market
price in order to maximize economics returns, but if high and volatile prices attract the most attention, low prices and volatility are problematic with extensive negative impacts on the agriculture sector, food security and the wider economy in both developed and developing countries \citep{TOSCANO2012}. For instance, a drought event in an area at risk seriously damage crops \citep{Dono2013,DONO2016,2016Vignaroli}. 

Furthermore, \cite{Khoury2014} have shown in their study that since 1961, human diets around the world have been changing and they have been becoming more similar. Those diets are composed for the most of few staple commodity crops, which ``have increased substantially in the share of the total food energy (calories), protein, fat, and food weight that they provide to the world's human population, including wheat, rice, sugar, maize, soybean (by +284\%), palm oil (by +173\%), and sunflower (by +246\%)''. 

This fact, on one hand has relieved the under-nutrition condition of poorest people but on the other has increased the dependence of worldwide supply of food on other factors, such as speculation, meteorological conditions either directly since it makes agriculture more vulnerable to major threats like drought or indirectly, favoring the spread of  insect pests and diseases, oil price volatility and the utilization of great portion of land to grow maize (corn) for biofuels production.  

The effects of relevant shocks such as the impressive sequence of fires in Russian Federation that reduced dramatically the production of the grain-growing and determined a peak of cereals price \citep{Welton2011} in the spring-summer 2011, provide a significant example of this dependence. \label{par:russia}

Even climate change and its perception play a role \citep{Nguyen2016}. \cite{Asseng2014-qa}, for example, demonstrate that rising temperature (+2$^\circ$ or +4$^\circ$) reduces wheat production at global scale, although with different local rates. Thus, price volatility is an undesirable feature of a market because it poses difficulties for both buyers and producers. 

These difficulties are amplified for wheat that, among other uses, is employed as a basic element of foods, nourishing a large share of the world population. Similarly to rice, maize and soybean, wheat has become a staple in over 97\% of countries \citep{Khoury2014}. Having a model able to understand how the staple commodity crops price is formed is thus of primary importance. To this aim we started building an Agent-based model (ABM) being convinced that this approach is the most suitable to account for the interaction among the several factors affecting the market price and to suggest stabilization policies.

In this paper, we build on an agent-based model which provides the basic elements to analyze commodities international markets. The work presented in this paper aims at adapting this model to a particular commodity, i.e. wheat,
in order to understand the worldwide wheat price formation and dynamics.
The paper is organized as follows.
We first present the functioning of the general version of the agent-based model.  We then explain the changes we made to this model to account for the peculiarities of wheat market. The following part of the paper describes the empirical data we have used to provide input to the model and the elaboration we have done to reach a coherent configuration of the model. 
The paper proceeds with a comparison of simulation outputs with empirical data which allows an assessment of the modeling choices and provide insights on how to integrate and develop the model. Finally, the results of the simulation for the case study of Russian Federation ban on wheat exports in 2010 are presented. Conclusions are drawn in the final section.   


\section{The CMS model}\label{par:begin_model}
The starting point of the work reported in this paper is the Commodity Markets Simulator (CMS). It is a computational model primarily addressed to the analysis of commodities spot price formation, its dynamics and the dynamics of exchanged quantities \citep{giulioni18}. Commodities have the common feature of being traded in international markets however each of them has special features such as seasonality in production, in demand and in their storable feature \citep{pirrong12}. In this section we describe the functioning of the generic version of the model which details can be found either in \citep{giulioni18} or in the software supporting material reporting the main features for the readers convenience (see the software Github repository at https://github.com/gfgprojects/cms). Details on how the model has been modified for analyzing wheat markets are given in paragraphs \ref{par:begin_sim_settings}-\ref{par:end_sim_settings}.

\subsection{Overview}\label{sec:1}

The Model has three types of agents: producers, buyers and markets.
These agents' common feature is that each of them has a geographic location given by a latitude and a longitude. 
In the most straightforward interpretation, Producers can be thought of as sovereign countries, but it is possible to setup the model thinking at different geographical scales such as continents, macro areas, or regions of a country. For convenience of exposition we will identify hereafter producers with countries.

Even though the model is fully customizable, we start describing a common setup for exposition convenience. The setup is as follows:
\begin{itemize}
	\item a producer has at least one associated buyer: if a country produces grains, it also uses the resource; 
	\item a buyer is not necessarily associated to a producer: to be more specific, the resource can be used by countries that does not produce it;
	\item the number of markets and their geographic location is independent from the number and location of producers and buyers.
\end{itemize}
Because we think the first two listed features straightforward, we will now focus on the third one.
%
%
\subsubsection{Market organization}
Considering nowadays information and communication technologies, we model markets as (virtual) places where producers and buyers send information. More trivially, resources are not physically moved to the market by the producer and, once sold, moved again from the market to the buyer. As it commonly happens, buyers and sellers send their will to the market. The market uses this information to reach an agreement. Once an agreement is reached, the resources are directly moved from the seller's to the buyer's place.

In this context, markets geographic location does not affect producers and sellers behavior. It has a role only when the model has more than one market: opening order is set according to their latitude.

Market are organized in sessions. Each market session is associated to a producer. The latter can have only one session in a market, and he must participate in at least one market. This organization allows buyers who bid in a given session to know who is the producer. The producer's geographic location has an important role here because it informs buyers on where the resource is stored. Because we assume buyers bear the transport costs, the proposed markets organization allows buyers to compute such costs and account for them when submitting their bids.
\subsubsection{Market Participants}
A producer can always decide to sell exclusively to its associated buyer. In the real world this happen when a country forbid export. Similarly, buyers who have an associated producers can decide to buy exclusively from their producer (a producer country can forbid import). The latter, is not possible if the considered buyer has not an associated producer (a non producer country does not forbid import). 
%
%
Summing up, in each market session participates 
\begin{itemize}
	\item the producers associated with the section;
	\item buyers associated with the producer;
	\item buyers not associated with the producer if the two following condition are both satisfied:
		\begin{itemize}
			\item the producer allows export;
			\item the buyer allows import
		\end{itemize}
\end{itemize}

\subsubsection{Dynamics}

The "cornerstone'' of the dynamics is the simulation time step. In each simulation time step, several actions can be performed however what mostly characterizes it, is that all the market sessions conclude the possible agreements.  
This provides a link between real and simulated time: if we want to simulate a real world situation where markets operate once a day (week, month and so on), a simulation step represents the same lapse. Starting from this observation we can comment the other simulation events. 
Consider for example the dynamics of the resource inventories. Straightforwardly, at each time step, each buyer's inventories are increased by the quantity bought in all the market sessions one participates, while each producer's inventories are decreased by the amount sold in the market sessions one is associated. Knowing the time scale is important to model the opposite flow. For buyers, the opposite flow to purchases is consumption. Therefore, if a simulation step represents a day, we have to take into account the daily consumption. Modeling the opposite flow is tricky for agriculture producers. The opposite flow to sales is the production flow. Agriculture products have not a continuous production, so we cannot compute a daily, weekly or monthly produced quantity as we can do for crude oil, for example. In general, agriculture production is harvested once a year. Our simulator accounts this issue giving the possibility to adapt the frequency of the production flow to the model time scale. Consider for example the situation with a monthly time step and an yearly production cycle. In this case, during the setup phase the researcher can (and must) choose to increase producers' inventories by the obtained production every 12 simulated time steps.
\label{par:dyn1}

%
%

We report hereafter the sequence of events that can happen in a simulation time step. It integrates and organizes the elements given above: \label{par:dynitems}

\begin{enumerate}
	\item producers decide if allow/forbid export
	\item buyers with an associated producer decide if allow/forbid import
	\item buyers update buying strategy
	\item market sessions are operated
		\begin{itemize}
			\item market 1
				\begin{itemize}
				\item session 1 
					\begin{itemize}
						\item producer sends supply curve
						\item buyers send demand curves
						\item demand curves are aggregated 
						\item market price and quantity are determined
						\item buyers increase inventories by the quantity bought in this session
						\item producer decrease inventories by the quantity sold in this session
					\end{itemize}
				\item all the other sessions (if exist, perform same actions listed for session 1)
				\end{itemize}
			\item  all other markets (if exist, perform same actions listed for market 1)
		\end{itemize}
	\item buyers account consumption
	\item producers produce
	\item producers update target production
\end{enumerate}

Following this list, below we detail the simulation loop.

\subsection{The model main loop}

\subsubsection{Producers decide if allow/forbid export}

Each producer has a boolean variable named \verb+exportAllowed+. 
This event consists in updating this variable with a true/false value.
The \verb+stepExportAllowedFlag()+ method of the \verb+Producer+ class can edited to modify producers' export behavior.
The frequency of this action ($\tau_{exp}$) can be changed setting the \verb+exportPolicyDecisionInterval+ parameter.

\subsubsection{Buyers with an associated producer decide if allow/forbid import}

Each buyer has a boolean variable named \verb+importAllowed+. 
This event consists in updating this variable with a true/false value.
The \verb+stepImportAllowedFlag()+ method of the \verb+Buyer+ class can be edited to modify buyers' export behavior.
The frequency of this action ($\tau_{imp}$) can be changed setting the \verb+exportPolicyDecisionInterval+ parameter.

\subsubsection{Buyers update buying strategy}\label{subsec:buyingStrategy}
Differently from the previous two actions, buying strategy is updated at each simulation time step.
Updating the buying strategy is an elaborate action. It is especially because buyers have to fulfill their needs looking for cheapest opportunities in a changing environment. The most relevant change in buyers' environment is represented by the possible switch in producers export policy. In fact, buyers cannot continue buying in sessions associated to those producers who change their policy forbidding export. In these cases, buyers attempt to collect elsewhere the quantities they were used to buy in these sessions. On the other hand, new opportunities open when producers switch their policy allowing export. 

Buyers' goals in updating their buying strategy are:
\begin{enumerate}
	\item reducing the unit cost;
	\item moving the demand usually directed to market sessions with recently imposed export restrictions to sessions currently open to international trade;
	\item designing demand curves for market sessions recently opened to international trade;
	\item obtaining the desired quantity.
\end{enumerate}

To understand how these tasks are implemented we specify that at each time step, each buyer send to the available sessions a demand curve.
In the current version of the model demand curves are assumed to be linear. The slope ($d_{b}$) can be different for each buyer,
but it is the same for the same buyer in the various market sessions. The buyer updates the buying strategy by managing the intercepts of the demand curves one will send to the various market sessions ($\bar{D}_{b,ms,t}$). The initial level of the demand intercepts is a parameter ($\bar{D}_{b,0}$).\label{par:barD}
Summarizing, the demand in market session $ms$ formulated by buyer $b$ at time $t$ is:
\[
D_{b,ms,t}=\bar{D}_{b,ms,t}-d_bp_{mc}
\]	
As an example, consider the situation represented in the following figure where a buyer participated (in the previous period) in three market sessions (labeled $A$, $B$ and $C$). \vspace{2mm} \\
\includegraphics{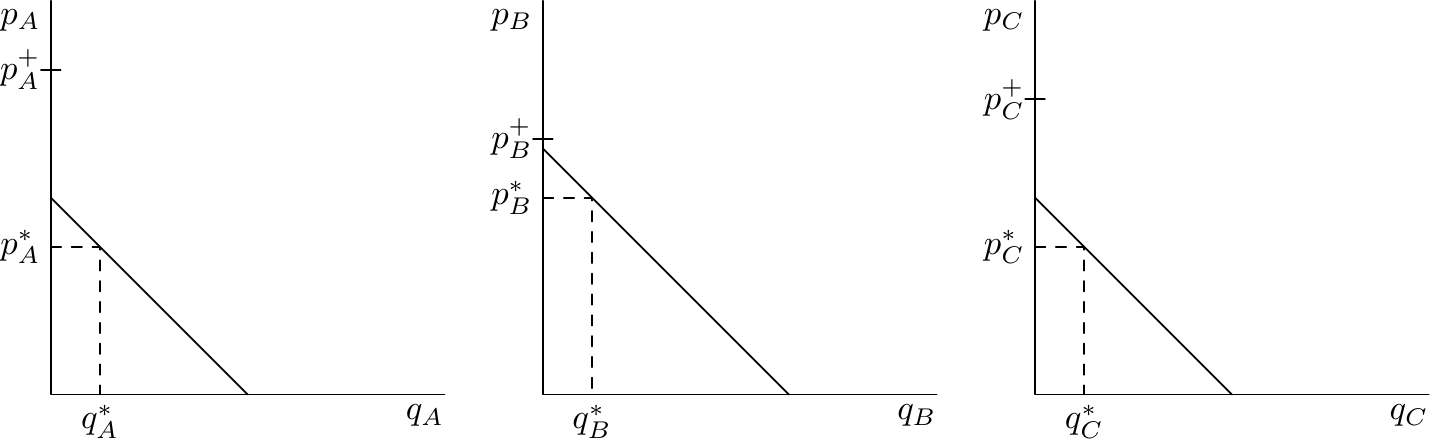} \vspace{2mm} 
The charts show that the market prices were $p^*_B>p^*_A=p^*_C$ and that the buyer bought the quantities $q^*_A$, $q^*_B$, $q^*_C$ and the total quantity $q^*_A+q^*_B+q^*_C$.

Consider however, that the transport cost ($c$) must be added to the market price ($p^*$) to obtain the unit cost of the commodity. Buyers use this cost, denoted $p^+$, to rank market sessions.   
Suppose now that the buyer we are considering and the producer selling in market session $C$ are from different countries and the latter forbids exports. So, the buyer have to gather the quantity $q^*_C$ in other available market sessions. The assumption is that the buyer performs this attempt in the market session with the lowest $p^+$. Looking at the chart above, we see that $p^+_B<p^+_A$. So, the demand curve in market session $B$ is shifted to the left by $q^*_C$ as happens in the following graphic representation. \label{par:transpc} \vspace{2mm} \\
\includegraphics{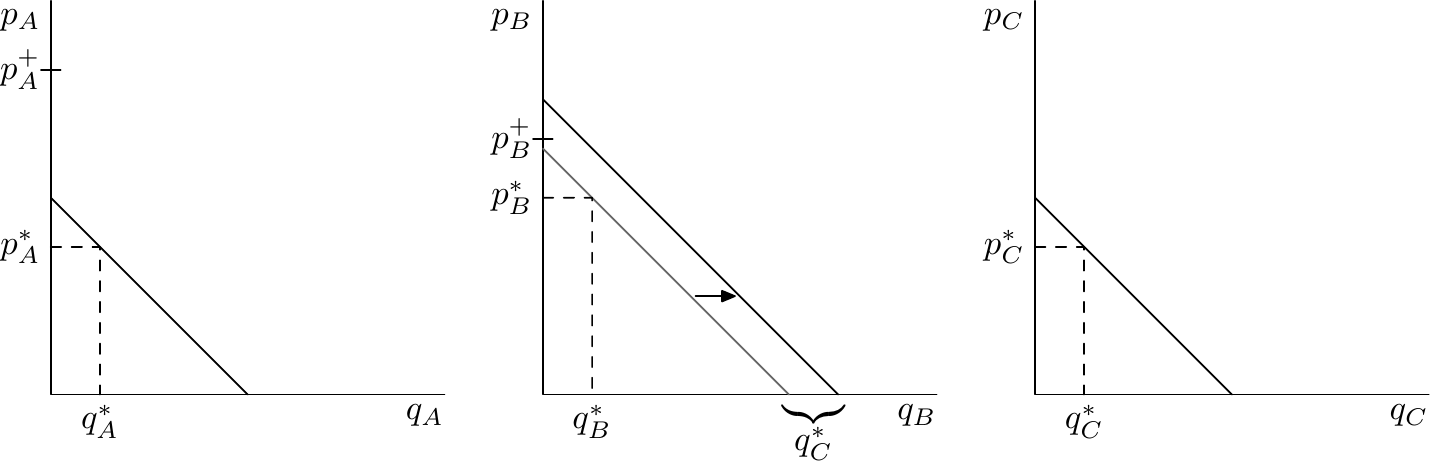} \vspace{2mm} \\
This behavior is operated to achieve goals 1 and 2 as listed above.

However, there will be no movements in the demand curves if the buyer can continue participating in all markets sessions. The attempt to reduce the unit cost (goal 1 in the list) is lost in this case. To allow for the achievement of this goal even when there are no export/import policy changes, we introduce the following device. The buyer can move demand from the most expensive market session ($A$ in our example) to the cheapest one ($B$). We introduce a tolerance to perform this action by using the parameter $\iota$. In particular, the movement is made if:\label{par:iota}
\[(1+\iota)p^+_{\min}<p^+_{\max}\]
the $\iota$ parameter corresponds to the \verb+toleranceInMovingDemand+ parameter in the code.

We furthermore introduce an additional parameter to control the size of this movement. Suppose the quantity bought in the market with the highest cost is $q_{p^+_{\max}}$. The quantity to be moved from the most expensive to the cheapest market ($q_{p^+_{\max} \rightarrow p^+_{\max}}$) can be tuned by setting the $m$ parameter in the equation:
\begin{equation}
	q_{p^+_{\max} \rightarrow p^+_{\max}}=mq_{p^+_{\max}}
	\label{eq:movedDemand}
\end{equation}
the $m$ parameter corresponds to the \verb+shareOfDemandToBeMoved+ parameter in the code.

The two charts below shows the leftward shift of the demand curve in market session $A$ and the rightward shift of the curve in market session $B$ after moving the quantity. \vspace{2mm} \\
\includegraphics{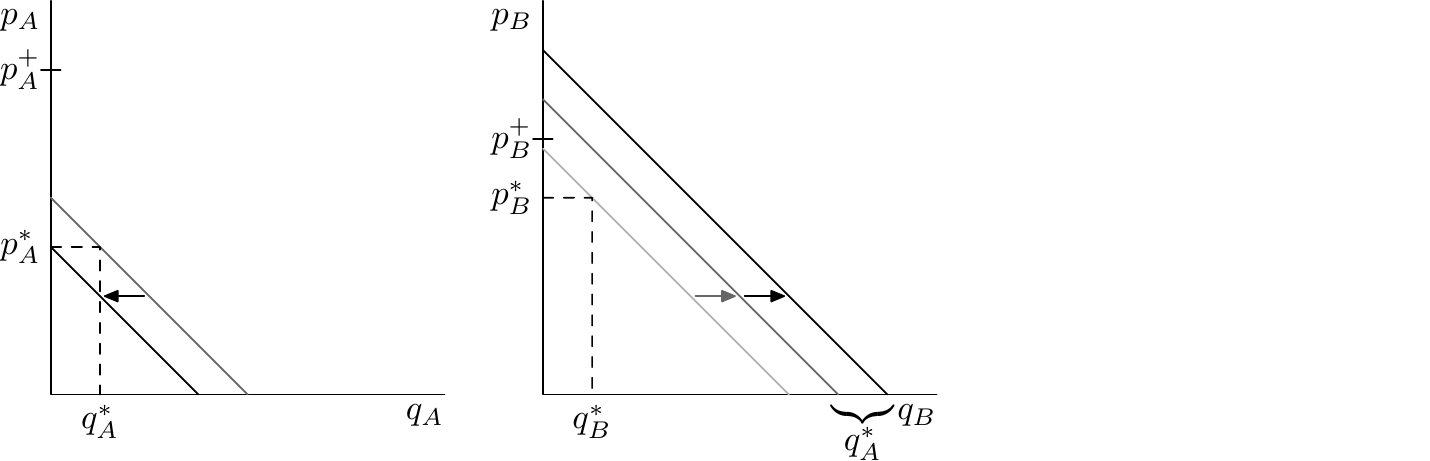}

Next, consider the case in which the buyer can participate in a new market session named $D$. The problem now is how to formulate the demand curve for this new session (goal 3). The idea is that the buyer is willing to buy in this new market only if the unit cost will be lower than the lowest unit cost observed in the previous period. In the previous figure, the lowest unit cost is $p^+_B$ while the transport cost for the new available producer is $c_D$. Thus the price in session $D$ should be lower than $p^+_B-c_D$. We add the parameter $d$ to tune the price mark down that convince a buyer to enter a new available market session. This allows to set the demand curve in the new market session as in the rightmost chart of the following figure.\label{par:demInNewMarket}\vspace{2mm} \\
\includegraphics{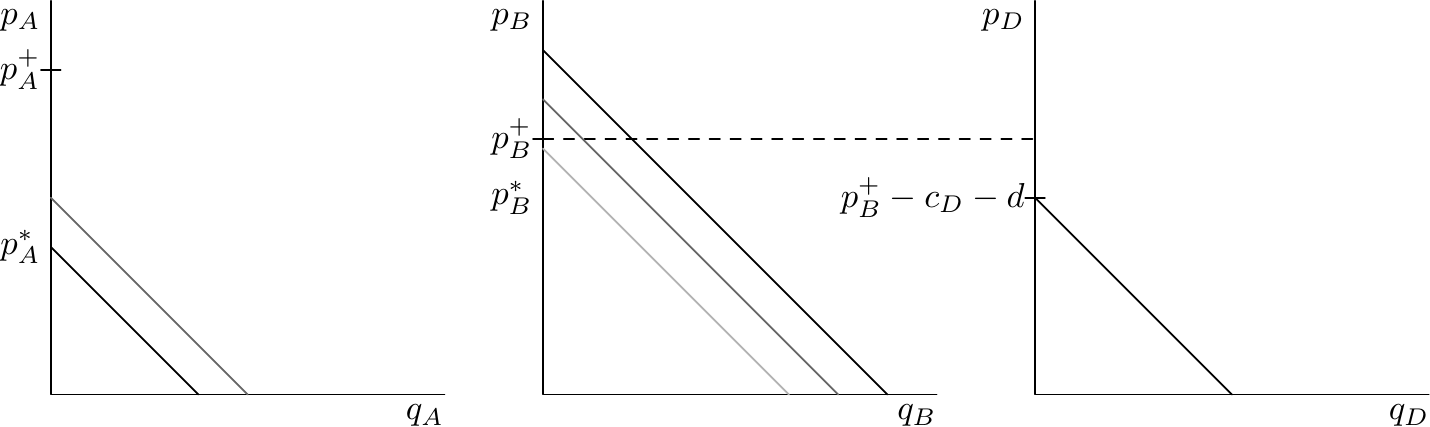}

The last movement of the demand curves before sending them to the market sessions is done when the quantity gathered in the market is low and people or authorities think this negatively impacts on the economy or on the population welfare. 

The reasoning starts from the definition of Buyers' minimum consumption ($C^{\min}_b$). It  is set at the beginning of the simulation as follows 
\[C^{\min}_b=c\frac{s_{b,0}P_0}{\tau}\]
where $0<c<1$, $s$ is the buyer market share, $P_0$ is the global production at initialization, and $\tau$ is the production cycle length.\label{par:tau} 
The division by $\tau$ deserves a comment. While the buying strategy is updated at each simulation time step, the global production relates to the production cycle length, this motivates the division by $\tau$. The following example should clarify it. Consider a situation where a simulation time step represents a month and the production is realized yearly. $P_0$ is the yearly global production and $s_{b,0}P_0$ is the amount bought yearly by the buyer. Since this event manages the monthly demand, we have to divide by 12 (which is the $\tau$ in this example). Now, having computed the average monthly consumption, we use it to set the lower bond of consumption by multiplying for the $c$ parameter.   

Consider now the buyer buys an amount $B_{b,t}$  in the considered time step (that in our example is $q^*_A+q^*_B+q^*_C$).
This quantity is then consumed and represents the buyer's effective consumption ($C_{b,t}$).

When the consumed quantity gets lower than the minimum consumption threshold ($C_{b,t}<C^{\min}_b$), people and authorities manage to obtain more commodity in the incoming period shifting the demand curve in the rightward direction.

All the demand curves are shifted by $\frac{\max(C^{\min}_b-C_{b,t},0)}{\#ms_{b,t+1}}$, where $\#ms_{b,t+1}$ denotes the number of market session the buyer will participate.

The following charts display the case in which the demanded quantities are increased because consumption was lower than the minimum. In fact, all the demand curve have a rightward shift. \vspace{2mm} \\
\includegraphics{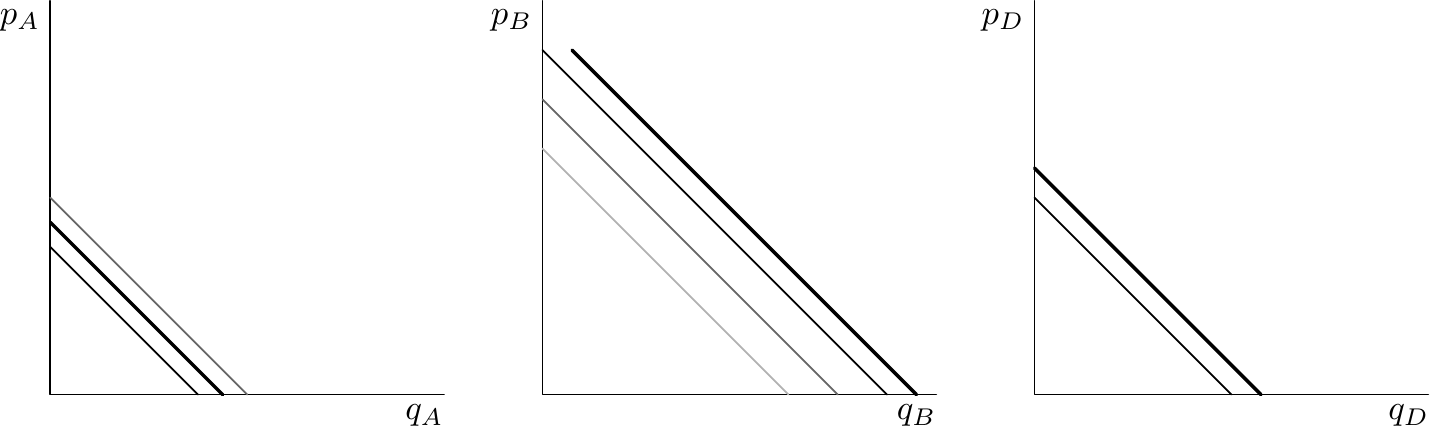}

Demand curves do not move if $C_{b,t}\ge C^{\min}_b$.

Because there are buyers without an associated producer, it may happen that there are no available session because all producers forbid exports. In these cases, if none of the producers switch the export policy, no demand curve is set, otherwise demand curves for newly available sessions are set. The idea is similar to that presented above however the buyer has not the benchmark of the cheapest unit cost at which s/he bought in the latest time step. Anyhow, the latest prices observed in the newly available market sessions are known. The current version of the model first sets the demand curve in each market in such a way that the demanded quantity is zero at the latest observed market price as displayed in the following charts. \vspace{2mm} \\
\includegraphics{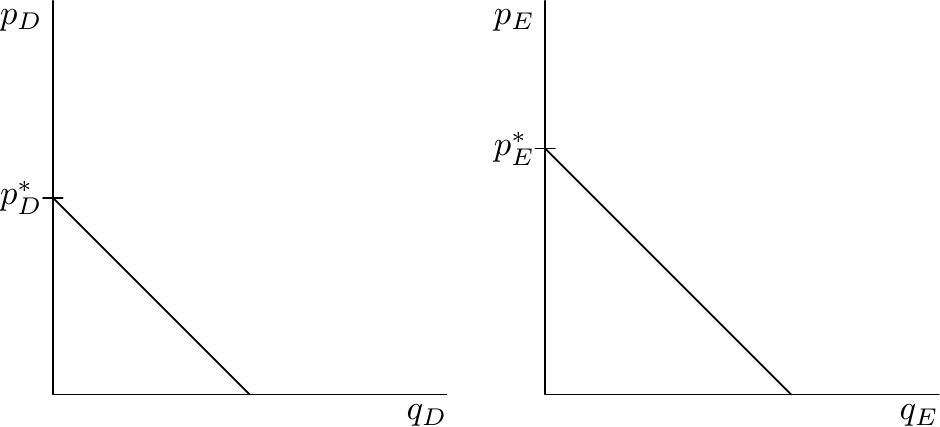}

Then, the demand functions are moved as explained above according to the $\frac{\max(C^{\min}_b-C_{b,t},0)}{\#ms_{b,t+1}}$ quantity.

Changes to the basic behaviors presented in this paragraph can be done modifying the method \verb+stepBuyingStrategy+ of the \verb+Buyer+ class.

As a last note on demand curves, we highlight a slight difference between demand directed to the domestic producer and these addressed to abroad producers. We introduce a parameter for tuning the minimum importable quantity ($I_{\min}$). If it is set to a positive value, quantities lower than this threshold are set to zero in the abroad demand curves. This correction does not applies for the domestic demand curve. 

\subsubsection{Perform market sessions}

We will now describe the functioning of a market session. 
First of all, we recall that in a market session, the exchanged items come from a given producer. This is basically because using the geographic location of producer, buyers can compute transport costs that are used to update the buying strategy as explained above.

We have already discussed how buyers set their demand curves, now we need to specify how the producer sets the supply curve. In the present version of the model, the easiest option of a vertical supply curve is adopted: the supplied quantity is independent of price. Despite this simplification, managing the supply policy is tricky when accounting for production that are not realized at every simulation time step and/or for producers who participates in more than one market session in each time step. 
Even in these cases, the simplest solution is adopted. At the beginning of each market session, the producer checks the level of the resource stock and divides it equally among the market sessions to come before the production is realized.   

Because in a market session there is one seller (producer), its supply curve represents the whole session supply curve. Differently, we can have more than one buyer attending a session (this is usually the case except when the seller forbids exports). When we have two or more buyers, the demand curves they send to the market are aggregated by summing them horizontally to obtain the session demand curve.

Now, using the session demand and supply curves, the market price and exchanged quantity are computed. The quantity bought by each buyer is obtained using the market price and the individual demand.
The following chart, which focuses on market session $A$, can better clarify it.

\begin{figure}[!h]
\centering
\centerline{\includegraphics{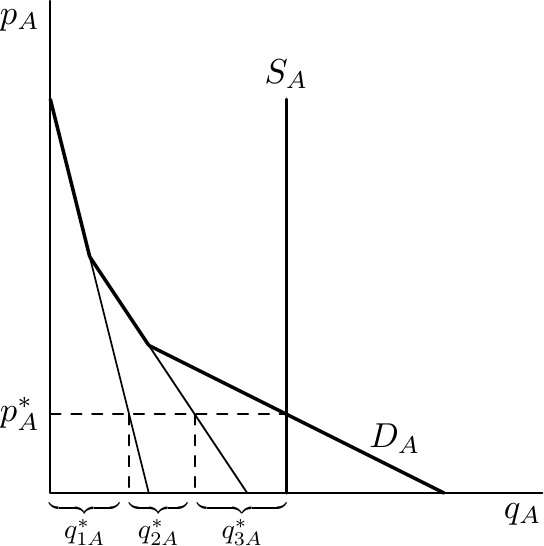}}
\caption{computation of market price and exchanged quantities}
\label{fig:market_equilibrium}
\end{figure}

Bold lines are the market session supply ($S_A$) and demand ($D_A$) curves. For exposition convenience it is assumed that there are three buyers in this market. The thin black lines keep track of the horizontal sum of the individual demand curves. The intersection point between the session demand and supply curves (bold lines) determine the market price ($p^*_A$) and the total exchanged quantity ($q^*_A$). The quantity bought by each buyer ($q^*_{1A}$, $q^*_{2A}$, $q^*_{3A}$) are also reported. Obviously $q^*_A=q^*_{1A}+q^*_{2A}+q^*_{3A}$.

At the end of the session agents update their inventories level.
For buyers we have:
\[ I_b=I_b+q_{bA}\]
while for the producer
\[ I_p=I_p-q^*_A\]

This updating is performed at the end of each market session, so, in a time step, the level of inventories is updated by each agent as many time as the number of session it participates.

\subsubsection{Buyers account consumption}

This action is performed to account for the consumption of resources occurred during the time step. Buyers' inventories are newly updated:
\[
I_b=I_b-C_b
\]
Being consumption a continuous phenomenon, this update is performed at every time step.

\subsubsection{Producers Produce}\label{subsubsec:producers}
This action is performed to account for the production of resources. Producers' inventories are newly updated:
\[
I_p=I_p+Y_p
\]
where $Y_{p}$ is the production realized in a period.

Differently from buyers, this update is not necessarily performed at each time step.
As mentioned above, there are commodities whose production is not continuous. For those commodities, if the time step represents a shorter time interval than the production one, this update is performed at regular interval.

\subsubsection{Producers Set Target Level of Production}\label{subsubsec:stock}
To evolve the target level of production ($Y^T$) the following parameters are used:
\begin{itemize}
	\item the producers prices memory length $ppml$;
	\item the high price threshold $p^h$;
	\item the low price threshold  $p^l$;
	\item the percentage change of target production $\delta Y^T$.
\end{itemize}
  
The software first computes the average price observed by producer $p$ in the latest $ppml$ market sessions, say $\bar{p}$, then the following rule is used to update $Y^T_p$:
\[
	Y^T_p=\left\{\begin{array}{l l l}
		Y^T_p(1+\delta Y^T)& if& \bar{p}>p^h\\
		Y^T_p(1-\delta Y^T)& if& \bar{p}<p^l\\
		Y^T_p& & otherwise\\
	\end{array}\right.
\]

The initial level of the production target $Y_{b,0}$ is computed as $s_pP_0$.
Then, it is updated at regular intervals applying the rule given above.

Note that this supply management effect can be disabled by setting $\delta Y^T=0$.

\section{The CMS-Wheat model}\label{sec:data}
\begin{figure}[!ht]
	\centering
	\includegraphics[scale=0.6]{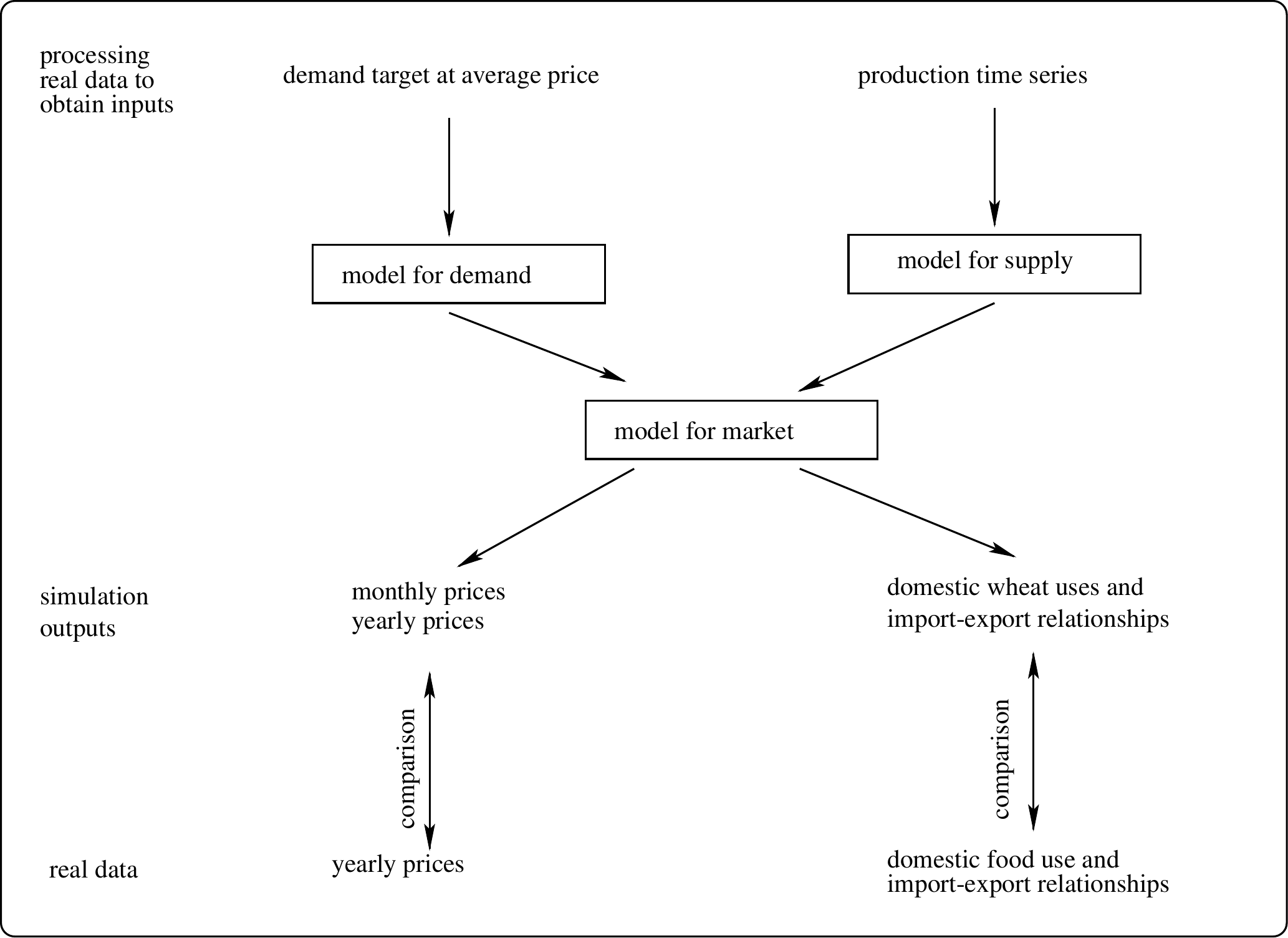}
	\caption{visual representation of the model}
	\label{fig:model_visual}
\end{figure}

The model presented above needs some specializations to analyze wheat.
We adopt a modeling strategy that provides for a gradual introduction of real world elements.
The comparison of simulation outputs with the corresponding real world data is a guide to progressively improve the modeling choices and to remove the shortcuts made to keep the initial versions of the model essential and easily understandable.       
Figure \ref{fig:model_visual} gives a visual representation of the model used in this paper. Following the figure flow, we describe hereafter its components: the real data used as inputs and as term of comparison for outputs, the modeling choices, and the comparison of the model output with empirical ones.  

\subsection{The data}

We use data at yearly time scale for 
\begin{enumerate}
	\item wheat production and uses,
%
%
	\item wheat prices 
	\item crude oil price.
\end{enumerate}
Items 1-2 of the list are from FAOSTAT dataset 
and  item 3 is from World Bank databases. Details on these data are given hereafter.

\begin{enumerate}
	\item 
 National and regional time series of yearly wheat production and uses were downloaded from \emph{Food Balance section} of FAO website (\href{http://www.fao.org/faostat/en/\#data/BC}{Commodity Balances - Crops Primary Equivalent}). 
This dataset contains time series of "Wheat and products", that includes Wheat; Flour wheat; Bran wheat;  Macaroni; Germ wheat; Bread; Bulgur; Pastry; Starch wheat;  Gluten wheat; Cereals breakfast; Wafers; Mixes and doughs; Food preparations, flour, malt extract. 
For production, the definition of "wheat" given in FAOSTAT  is as follows: "\textit{Triticum spp.: common (T. aestivum) durum (T. durum) spelt (T. spelta). Common and durum wheat are the main types. Among common wheat, the main varieties are spring and winter, hard and soft, and red and white. At the national level, different varieties should be reported separately, reflecting their different uses. Used mainly for human food}". Given this definition, we jointly model soft and hard wheat since we are not able to distinguish between one and another. 
The FAO wheat dataset contains a rich set of variables: 
Import, Export, Domestic supply quantity (Production + Imports - Exports + Changes in stocks), Food supply quantity, Stock Variation, Feed,  Other uses, Seed, Waste (tonnes).
These variables are the component of the wheat sources/uses balance equation on which the present paper relies on.

To understand the validation procedure implemented in this paper, some detail on the balance equation are given.

Formally, the balance equation is the following: 

\vskip4mm
$production + import - export + stock\ variation =$ 
\begin{equation}
 = Food+Feed+Seed+Other\ uses + processing + waste
	\label{eq:balance}
\end{equation}

Some specification on the involved variables might be useful.
\begin{itemize}
	\item 
The Production variable contained in this dataset is approximately the same as the one in "Crops section" of FAO Production dataset (\href{http://www.fao.org/faostat/en/\#data/QC}{Crops}). 
\item
According to FAOSTAT definitions, a negative sign in stock variation corresponds to an increase in stock. Stock variation is thus defined as \textit{initial stock - final stock}. 
\item
Statistical discrepancy makes the aggregation of balances equation across world regions and countries not exact (see appendix, paragraph \ref{par:world_unbalance} for the details).
\end{itemize}






\item
Yearly time series of wheat prices for most of the producing countries are used to validate the model output. An aggregated time series of yearly wheat price has been calculated averaging prices of selected producers by means of weighted arithmetic mean, where production volume is used as weight. 

Because they will be compared with our simulation results, the most important prices time series are reported in Figure \ref{fig:observed_prices}.

\begin{figure}[!ht]
\centering
\centerline{\includegraphics{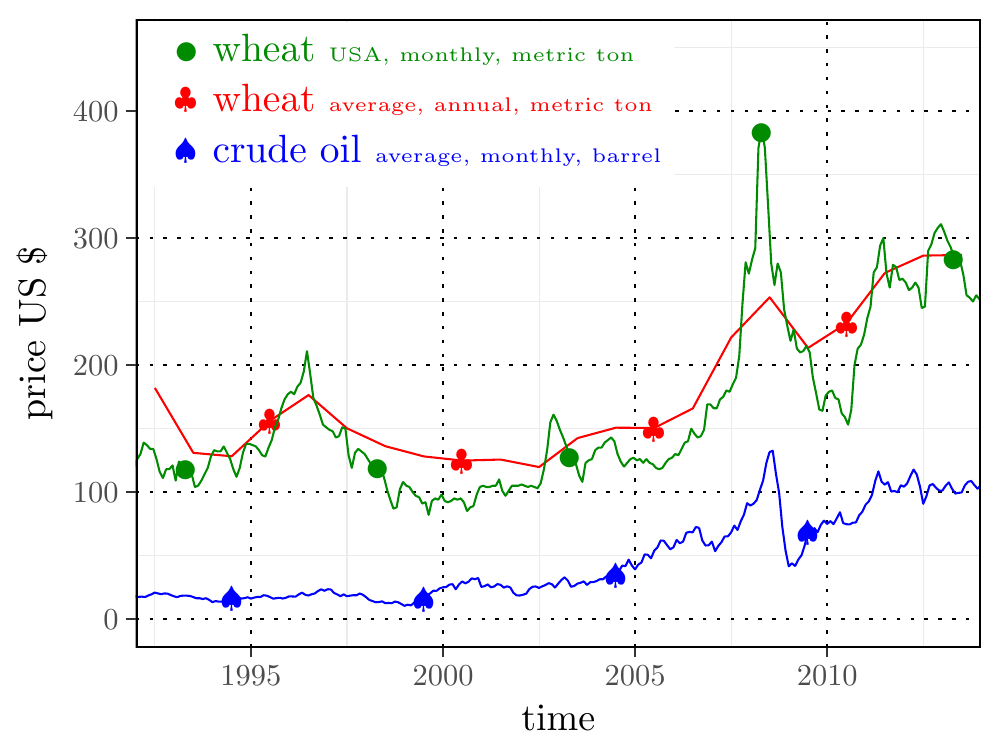}}
\caption{Relevant prices}
\label{fig:observed_prices}
\end{figure}

\item
Crude oil price is from World Bank Global Economic Monitor (GEM) Commodities database. The item has the following  
description: Crude Oil (petroleum), simple average of three spot prices; Dated Brent, West Texas Intermediate, and the Dubai Fateh, US Dollars per Barrel.
Data are available in the \href{http://databank.worldbank.org/data/reports.aspx?source=global-economic-monitor-(gem)-commodities}{World Bank} website.
The dynamics of crude oil price is displayed in figure \ref{fig:observed_prices} together with those of prices.
\end{enumerate}

%
%


Below we will proceed to aggregations and decompositions of quantity data. It is therefore useful to give information on this topic.   
Table \ref{tab:top20} reports data on the top 20 worldwide wheat producers. Countries were ranked according to averaged production over the period 1992-2013.
\begin{table}[!h]
\centering
\begin{tiny}
\begin{tabular}[t]{l|l|l|l|l}
\hline
Country & Production (tons.) & World quote (\%) & Yield (hg/ha) & Harvested Area (ha)\\
\hline
China, mainland & 107553131 & 17.4 & 42086 & 25824041\\
\hline
India & 73447196 & 11.9 & 27135 & 26927187\\
\hline
United States of America & 59837235 & 9.7 & 27908 & 21545498\\
\hline
Russian Federation & 44150970 & 7.2 & 18994 & 23017824\\
\hline
France & 35644793 & 5.8 & 68948 & 5167419\\
\hline
Canada & 25662926 & 4.2 & 25171 & 10271374\\
\hline
Germany & 21493799 & 3.5 & 73073 & 2929144\\
\hline
Australia & 20491377 & 3.3 & 17318 & 11782063\\
\hline
Pakistan & 20197472 & 3.3 & 23860 & 8431478\\
\hline
Turkey & 19623609 & 3.2 & 22346 & 8849862\\
\hline
Ukraine & 17384717 & 2.8 & 28902 & 5903731\\
\hline
United Kingdom & 14614000 & 2.4 & 76715 & 1903574\\
\hline
Argentina & 12798788 & 2.1 & 24752 & 5204802\\
\hline
Kazakhstan & 11906664 & 1.9 & 9883 & 11943083\\
\hline
Iran (Islamic Republic of) & 11256451 & 1.8 & 17610 & 6384640\\
\hline
Poland & 8865031 & 1.4 & 37513 & 2373448\\
\hline
Italy & 7542812 & 1.2 & 34754 & 2185268\\
\hline
Egypt & 6978899 & 1.1 & 61260 & 1130640\\
\hline
Spain & 5662888 & 0.9 & 27102 & 2101009\\
\hline
Romania & 5687618 & 0.9 & 27080 & 2070757\\
\hline
\end{tabular}\end{tiny}
\caption{Top 20 worldwide wheat producers (averaged values 1992-2013)}
\label{tab:top20}
\end{table}
The table reports the country percentage of worldwide production; the correspondent averaged yield, i.e. hectogram of production per hectare; the averaged portion of land utilized to grow wheat (Harvested Area). The wheat production of Top 20 and Top 5 amount to \(86.1\%\) and \(52\%\) of worldwide production, respectively. 



\subsection{Simulation settings}\label{par:begin_sim_settings}

In this section we retrace the description of the general version of the model to discuss the customizations and the modeling choices made in order to specialize the model for wheat.

\subsubsection{Agents}
We set up the simulation to obtain a level of aggregation suitable to investigate international prices formation and exchanged quantities.
In general, we use FAOSTAT regions which are sub-continental geographic areas gathering several countries. 
However, when a region includes a country (countries) playing a relevant role in the wheat world production/consumption system, we further partition the region to treat the important country (countries) as individual entities (see Appendix, paragraph \ref{par:Aregions} for more details). 
At the end of this process we end up with the geographic areas reported in Table \ref{tab:zoneslist}.

\begin{table}[!h]
	\centering
	\begin{tiny}
	\begin{tabular}{l l l l}
		\hline
Continent & Region& Has an international market? & incoming hub\\
\hline
\hline
\multirow{5}{*}{Africa} & Eastern&no&Ethiopia\\
\cline{2-4}
 & Middle&no&Angola\\
\cline{2-4}
 & Northern&no&Egypt\\
\cline{2-4}
 & Southern&no&South Africa\\
\cline{2-4}
 & Western&no&Nigeria\\
\hline
\multirow{5}{*}{America} & Northern - USA&yes - USA&USA\\
 & Northern except USA&yes - Canada& Canada\\
\cline{2-4}
& South &yes - Argentina&Brazil\\
\cline{2-4}
& Central&no&Mexico\\
\cline{2-4}
& Caribbean&no&Cuba\\
\hline
\multirow{9}{*}{Asia} & Southern - India&yes - India&India\\
 & Southern - Pakistan&yes - Pakistan&Pakistan\\
 & Southern except India \& Pakistan&no&Iran\\
\cline{2-4}
 & Central - Russian Federation&yes - Russian Federation&Russian Federation\\
 & Central except Russian Federation &yes - Kazakhstan&Uzbekistan\\
\cline{2-4}
 & Eastern - China&yes - China&China\\
 & Eastern except China&no&Japan\\
\cline{2-4}
 & South-Eastern&no&Indonesia\\
\cline{2-4}
 & Western &no&Iraq\\
\hline
\multirow{4}{*}{Europe} & Eastern&yes - Ukraine&Poland\\
\cline{2-4}
 & Northern&yes - United Kingdom&United Kingdom\\
\cline{2-4}
 & Western&yes - France&Netherland\\
\cline{2-4}
 & Southern&no&Italy\\
\hline
Oceania& &yes - Australia&New Zeland\\
\hline
	\end{tabular}\end{tiny}
	\caption{Geographic regions and markets}
	\label{tab:zoneslist}
\end{table}

\begin{figure}[ht]
	\centering
	\includegraphics[width=12cm]{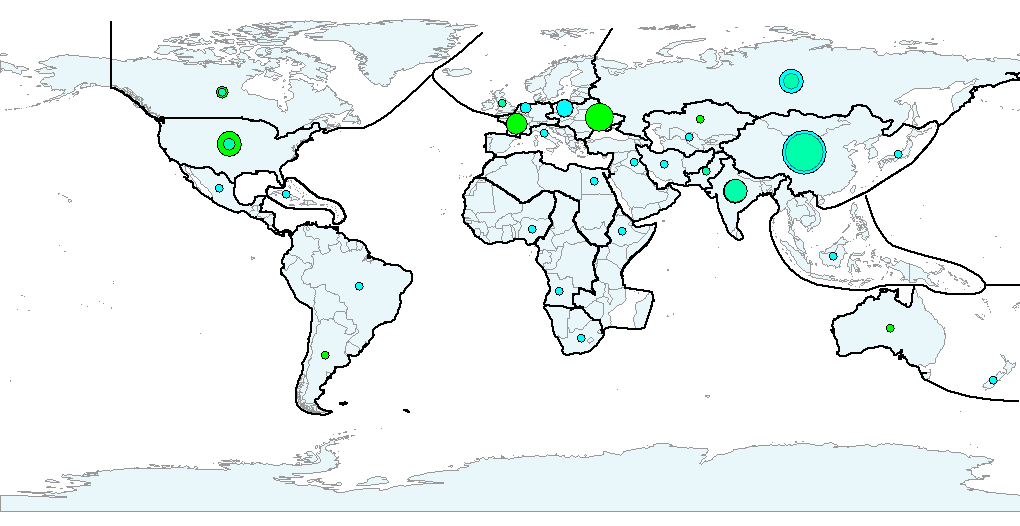}
	\vskip-5mm
	\caption{Commercial hubs considered in the model. Green circles denotes outgoing hubs, light blue circles denotes incoming hubs. The size of the circles is associated to produced and used quantities. The tick line delineates the regions boundaries.}
	\label{fig:hubs}
\end{figure}

Almost all the regions realize a wheat production but only a few of them are relevant at international level. To retain only internationally relevant producers in our analysis and keeping world demand-supply balance, we proceed as follows. 
The net demand was computed for each region as the difference between wheat demand and supply. Regions having a positive net demand in all the years of the time span considered were assumed to consume their production internally. Their production was therefore set to zero and their demand was replaced by their net demand.   
As a result of this process, the artificial wheat world trade system considered in this study is populated by 12\footnote{Note that the 12 outgoing hubs are in the top 14 worldwide wheat producers listed in table \ref{tab:top20}. The two missed top 14 producers are Germany and Turkey. Germany was excluded because it is second to France in the Western Europe regions. Turkey was not considered because it belongs to the Western Asia region which is a net importer and therefore has not an outgoing hub.} sellers (these having a ``yes'' in the  "Has an international market?" column in Table \ref{tab:zoneslist}) and 24 buyers.  
\label{par:producers_reduction}
For each of these geographic areas, the most important commercial hubs were identified. Again, in Table \ref{tab:zoneslist} the column "Has an international market?" highlights the market location (see also Table \ref{tab:top20} for a complete ranking of these producers) and the "Incoming hubs" column reports the top importer of each zone. The position of outgoing and incoming hubs are shown in Figures \ref{fig:hubs}, which also gives information on quantities offered and used by each region. 

Buyers and sellers interact in a  single global market having 12 sessions (one for each producer).

\subsubsection{Dynamics}
The time window for the simulation is set to the period 1992-2013 and the model run on a monthly base. This means that in a simulation time step all the market sessions are performed while each producer harvest every 12-time steps (see paragraph \ref{par:dyn1} for details). Harvesting time is asynchronous and mirrors reality. 

According to the dynamics of events reported in paragraph \ref{par:dynitems} the model gives the possibility to change import/export policies (items 1 and 2). As we will highlight below in the text we will use this features to investigate the effect of the 2010 Russian Federation export ban on international wheat price. Except for this, the model implements a completely open global market environment. In these settings, transport costs are the decisive factor that relates the demand function a buyer sends to a market session with the distance from the producer selling in the considered session.
Transport costs per unit of product ($c$) bear by buyer A to carry home resource bought from producer B are modeled as follows:
\begin{equation}
c_{A,B}=a\ kkm_{A,B}+b\ O_p\ kkm_{A,B}
	\label{eq:transportCosts}
\end{equation}
where $O_p$ is the oil price, and $kkm_{A,B}$ is the distance in thousand kilometers. $a$ is a fix cost per thousand kilometer and $b$ is the oil needed to carry one unit of product for thousand kilometer. 

\subsubsection{Demand}

\begin{figure}[htb]
	\begin{minipage}[t]{7cm}
	\centering
	\includegraphics{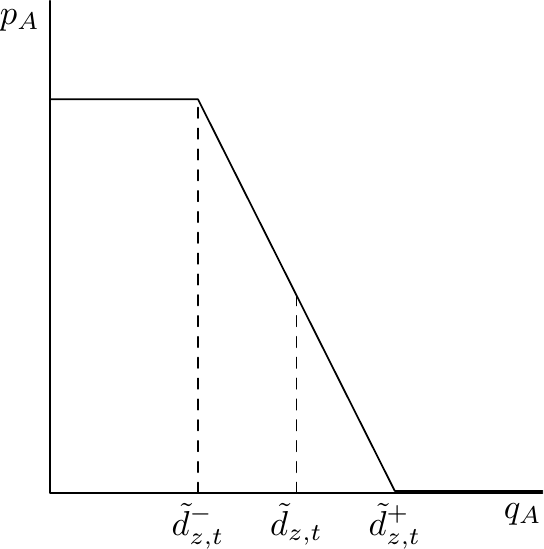}
	\caption{region $z$ demand curve for wheat produced by region $A$.}
	\label{fig:demand_curve}
\end{minipage}
\begin{minipage}[t]{7cm}
	\centering
	\includegraphics{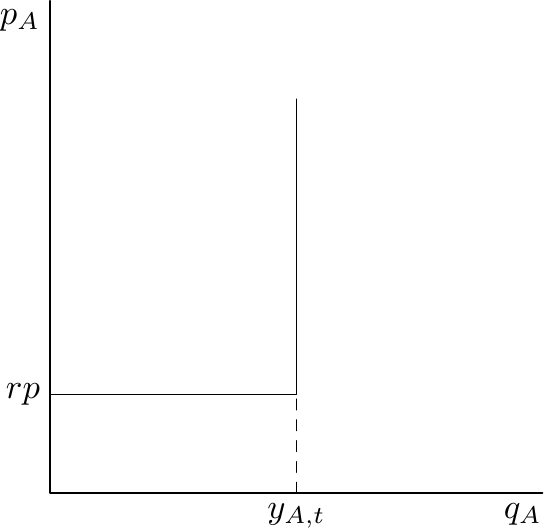}
	\caption{region $A$ supply curve}
	\label{fig:supply_curve}
\end{minipage}
\end{figure}

As explained in paragraph \ref{par:transpc}, the final product unit cost is $p^+_{A,B}=p^*_B+c_{A,B}$. 
Buyers compute this value for each market and move a share of demand from the highest to the lowest $p^+$. This brings us to discuss the changes made to the demand curve shape.  
We use linear demand curves as mentioned above. In initializing the demand curves position we account for the size of the buyer and that of the producer in order to avoid big countries address too large demands to small producers and vice versa. In this way we set the target level of demand ($\tilde{d}$) which is the quantity demanded at the average price level (see figure \ref{fig:demand_curve}).\footnote{For computational convenience, demand curves are defined in a given price range, namely 0-10, therefore the average level of price is 5. Given this simplification, we will rescale both simulated and real prices in order to make comparison as will be clarified in the results section.} We then set the slope of the demand function in such a way that the demanded quantity increases by a given percentage (the parameter $\delta_D$) when the price equals zero. Therefore, the demand curve is a straight line going from $\tilde{d}^-:=\tilde{d}(1-\delta_D)$ to $\tilde{d}^+:=\tilde{d}(1+\delta_D)$ as displayed in Figure \ref{fig:demand_curve}. 
Furthermore, there is a level of price $\bar{p}_z$ above which the wheat is out of range because too expensive for the country. It is straightforward to think this threshold is heterogeneous across countries, with poor countries having a low level of this price. However, we take it homogeneous in order to keep the model simple.\label{par:deltaD} 

Demand curves continuously move in time to allow buyers geographic regions to gather the desired quantities at the lowest price.

Desired quantities ($\tilde{d}$) are inputs for our agent-based model. Comprehensibly, they are not included in any database. We therefore decided to infer them from the FAOSTAT data using a calibration procedure which is better described below and in the appendix.

\label{par:steep_flat}

\subsubsection{Supply}

In order to keep the wheat variant of the model simple and in line with the general version presented above, we use a vertical supply curve. In addition to the basic model, a reservation price below which the producer is not willing to sell was introduced. The reservation price ($rp$) is given by:
\begin{equation}
	rp=a_{rp}+b_{rp}\ O_p
	\label{eq:reservationPrice}
\end{equation}
This functional form is based on a linear unit production cost composed of a fix part and a second component proportional to oil price. 

A second amendment to the supply curve of the original model involves the rule used to update the monthly offered quantity. 
We let producers distribute the harvested quantity uniformly on the twelve market sessions. 

Figure \ref{fig:supply_curve} shows the supply curve of a producer region (say region $A$) which is offering wheat at the international level. As for demand curves, the supply curves change in each time step. In particular, the horizontal portion moves up or down at each time step (i.e. monthly) according to the oil price level, while the monthly offered quantity ($y_A$), i.e. the vertical portion, moves left or right after the harvesting month (i.e. at every 12-time steps) and keeps still for the rest of the year.  
\label{par:end_sim_settings}

\subsubsection{Market equilibrium and disequilibrium}

\begin{figure}[!h]
\centering
\centerline{\includegraphics{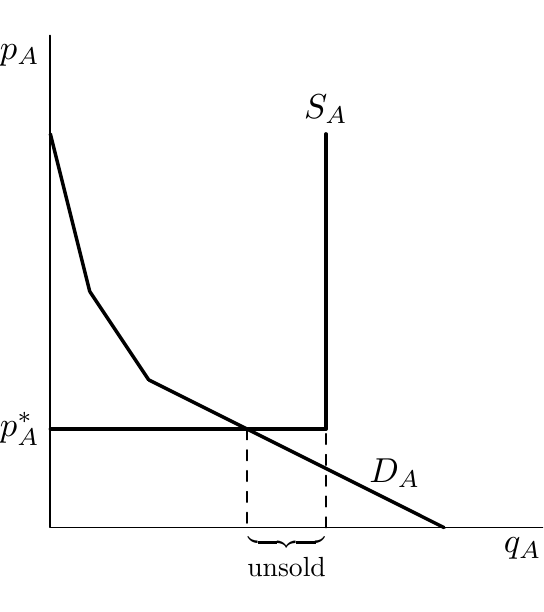}}
\caption{market in disequilibrium due to a demand fall or production boom.}
\label{fig:market_disequilibrium}
\end{figure}

The modification to the demand and supply curves just described brings the possibility to observe market disequilibrium even in the centralized market structure used in this model. A market is in equilibrium when the intersection point belongs to the downward section of the demand curve and to the vertical section of the supply curve. This situation is basically the one already seen in figure \ref{fig:market_equilibrium}. One possibility of disequilibrium is characterized by intersection points belonging to the horizontal section of the supply curve. This happens when the demand is too low compared with the offered quantities. Because this is a situation observed in our simulations, we report it in figure \ref{fig:market_disequilibrium}. In this case, as displayed in the figure, producers countries does not succeed in selling the whole quantity offered in the market. In another form of market disequilibrium buyers leave the market without the quantity they wish. If heterogeneous $\bar{p}_z$ are considered, the market demand curves have horizontal sections (especially at high prices). Buyers rationing happens if the intersection point falls in one of these portions. Furthermore, it can happen that the market price is higher than some countries reservation prices. Therefore, these countries do not buy wheat at all. However, our simplification of homogeneous buyers reservation price excludes the realization of this case in our simulations. Other forms of market disequilibrium, such as those due to decentralized market structure, cannot either be observed in our simulation. These two cases represent opportunities for future developments of the model.  

\subsection{Calibration procedure}

The model has several parameters. Setting some of them is straightforward or can be done using economic observation. A calibration procedure has been implemented to set those of them for which no other way is feasible. Table \ref{tab:parameters} reports the parameters. It also signals with * the parameters that was calibrated using the differential evolution algorithm. 

\begin{table}[!ht]
	\centering
	\begin{tiny}
	\begin{tabular}{l l l l} 
		\hline
		parameter & notation & value & reference \\
		\hline
		\hline
		production cycle & $\tau$ & 12 & in paragraph \ref{par:tau}\\
		tolerance in moving demand & $\iota$ &0 & in paragraph \ref{par:iota}\\
		fix cost in reservation price & $a_{rp}$ & 1 & in equation \ref{eq:reservationPrice} \\
		slope in reservation price & $b_{rp}$ & 0.02 & in equation \ref{eq:reservationPrice} \\
		transportCostsTuner slope & $b$& 0.01 &in equation \ref{eq:transportCosts}\\
		\hline
		transportCostsTuner intercept* & $a$&0.05 &in equation \ref{eq:transportCosts}\\
		shareOfDemandToBeMoved* & $m$ & 0.1 & in equation \ref{eq:movedDemand}\\
		percentageOfPriceMarkDownInNewlyAccessibleMarkets*&$d$&0.05& in paragraph \ref{par:demInNewMarket}\\
		demandFunctionInterceptTuner* &$\bar{D}_{b,0}$&0.5&in paragraph \ref{par:barD}\\
		demandFunctionSlopeTuner* &$\delta_D$&15&  in paragraph \ref{par:deltaD}\\
		\hline
	\end{tabular} \end{tiny}
	\caption{Parameters setting (symbol * signals the using of differential evolution algorithm for the calibration).}
	\label{tab:parameters}
\end{table}

In addition, as hinted above and  better explained in the appendix, we commit to infer the desired demand of each region starting from the observed level of exchange found in the FAOSTAT dataset using the following equation
\[\tilde{D}_{z,t}=(1-\tilde{\eta}^d_t)D_{z,t}\]
we therefore need to calibrate the $\tilde{\eta}^d_t$.

Parameter calibration is driven by the objective of yearly price replication. Two different techniques are used for $\tilde{\eta}^d_t$s and for $a,\ m,\ d,\ \bar{D}_{b,0}$ and $\delta_D$. These two techniques are nested in a recursive procedure. The $\tilde{\eta}^d_t$s are initially set to 1 and the last five parameters reported in Table \ref{tab:parameters} are set by running the differential evolution algorithm. Once these parameters have been set, a new configuration of $\tilde{\eta}^d_t$s is searched by running an algorithm inspired by the gradient method. The procedure is as follows. After a model run, yearly prices $\tilde{p}_t$ are computed. If the price from simulation output is higher than the real price in a given year, the $\tilde{\eta}^d$ for that year is increased. $\tilde{\eta}^d$ are simultaneously changed according to the following equation:
\[
	\tilde{\eta}^d=\tilde{\eta}^d+\left(\frac{1}{1+e^{-\beta(p_t-\tilde{p}_t)}}\right)0.02-0.01
\]

Once this process has completed, a new round is started running again the differential evolution.

We use this nested process to speed up the parameter estimation. Indeed, a joint calibration of all the parameters using the differential evolution algorithm is possible, but it would be extremely demanding from the computational point of view. Using the gradient inspired method for the $\tilde{\eta}^d$ significantly speed up the process.  

\subsection{Comparing simulated and empirical data}\label{sec:results}

The model outputs the 12 producers' monthly prices and the exchanged quantities. The latter is especially interesting in order to identify each buyer wheat availability and the origin of these quantities.   
Result on prices and quantities are shown in next subsections.

\subsubsection{Prices}
\begin{figure}[ht]
	\centering
	\includegraphics[scale=1.0]{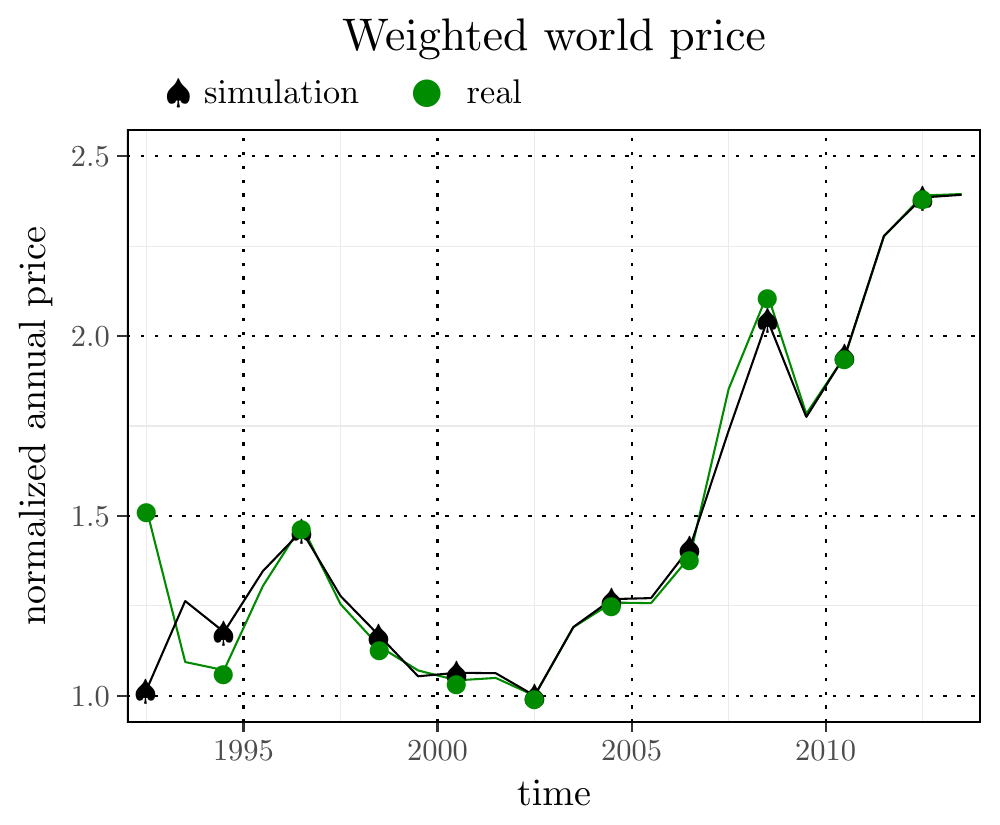}
	\caption{Yearly wheat price time series from real data and from a simulation with calibrated parameters.}
	\label{fig:sim_vs_real_annual}
\end{figure}

The comparison between simulation output and real world data is straightforward for yearly prices because they are fully available for each country. Figure \ref{fig:sim_vs_real_annual} compares simulation outputs with the weighted average price of the 12 producer countries. The weight is given by the share sold by the country with respect to the sum of the total quantity sold. As already mentioned, both prices and sold quantities are available in real world data at yearly frequency. Simulations provides instead monthly values. Therefore, yearly aggregation is computed from simulation data by first calculating the monthly weighted prices and then averaging them every 12 periods. Since simulation unit measure for prices is different from real data one, we normalize the values. Figure \ref{fig:sim_vs_real_annual} shows the yearly weighted world price fit. 
The prices observed in real data are accurately reproduced by our model. Although this is a consequence of the calibration procedure described above which has the objective of minimizing the distance between simulated and real prices, this result signals that the model grasps the essential elements of the wheat international exchanges.   

\subsubsection{Quantities}

By using simulation output, we can compute the quantity bought in the domestic market session, the quantities bought by each of the other producers, and those sold to each buyer. 
It is thus possible to compute the import and export time series on a monthly base and obtain the corresponding annual series by summing over every 12 periods. 

The model simulation output tracks for each area the quantity bought from all open markets. We have therefore to connect these quantities with the variables in the FAOSTAT database. We already described in equation \ref{eq:balance} the relationship among the variables included in the database.
We will use the following notation:\\
$fo_{tot}$ food;\\
$fe_{tot}$ feed;\\
$se_{tot}$ seed;\\
$opw_{tot}$ other uses+processing+waste.

The right hand side of the balance equation gives us a proxy of the wheat used (thus bought) by a country in each year regardless of its origin (domestic of foreign). In this version of the model, simulations do not provide specific results of those categories, i.e. food, feed, seed, and other uses but they supply the domestic or foreign source of bought quantity: $q_{dom}$ and $q_{imp}$, respectively. To compare observed and simulated bought quantity, we observe that each term in the right hand side mixes domestic and foreign wheat, so that we can write:
\[fo_{tot} + fe_{tot} + se_{tot} + opw_{tot}=\underbrace{fo_{dom}+fo_{imp}}_{fo_{tot}} + \underbrace{fe_{dom}+fe_{imp}}_{fe_{tot}} + \underbrace{se_{dom}+se_{imp}}_{se_{tot}} + \underbrace{opw_{dom}+opw_{imp}}_{opw_{tot}}=\]
\[=\underbrace{fo_{dom} + fe_{dom} + se_{dom} + opw_{dom}}_{q_{dom}}+\underbrace{fo_{imp} + fe_{imp} + se_{imp} + opw_{imp}}_{q_{imp}}=q_{dom}+q_{imp}\]
These observations allow us to compare the quantity $fo_{tot} + fe_{tot} + se_{tot} + opw_{tot}$ obtained from FAOSTAT database with the quantity $q_{dom}+q_{imp}$ obtained from our simulations.

Recall that starting from the right hand side of the balance equation, we compute the target level of demand $\tilde{D}$ which is given to the model as an input.

\begin{figure}[ht]
	\centering
    \includegraphics[scale=1.0]{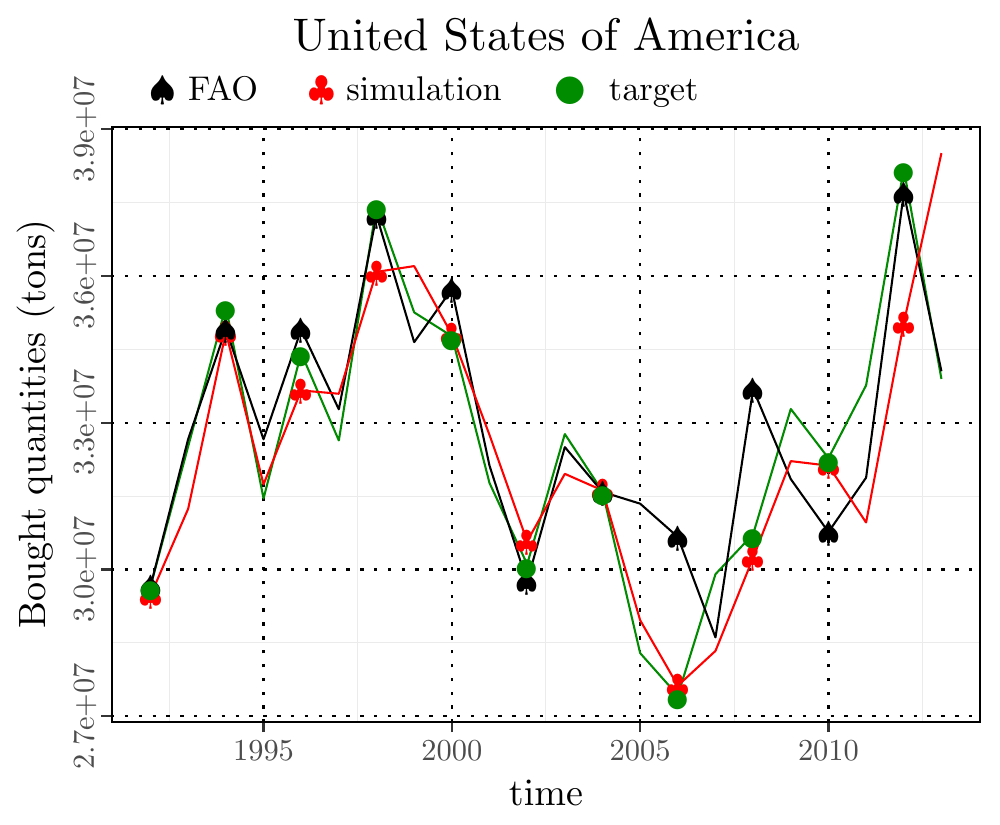}
	\caption{Comparison of USA used quantities in real data (FAO), in simulations and target, i.e. the USA desired demand at average price (yearly time series).}
	\label{fig:used_USA}
\end{figure}

The comparison of these three variables (real, simulation output and simulation input) provides an opportunity to evaluate the performance of the model by the quantity side.
Figure \ref{fig:used_USA} shows the dynamics of these three variables for the United States of America to provide an example. The charts for all the other regions can be viewed at \url{http://erre.unich.it/wheat\_map/bought_png.html}.   

Although the replication of quantities has no weight in the calibration process, the dynamics of simulated quantities is compatible with those observed. However, we expect possible improvements in the future by giving weight to quantities in the calibration objective function.

\begin{figure}[htb]
	\begin{minipage}[t]{7.5cm}
	\centering
		\includegraphics[width=10.5cm]{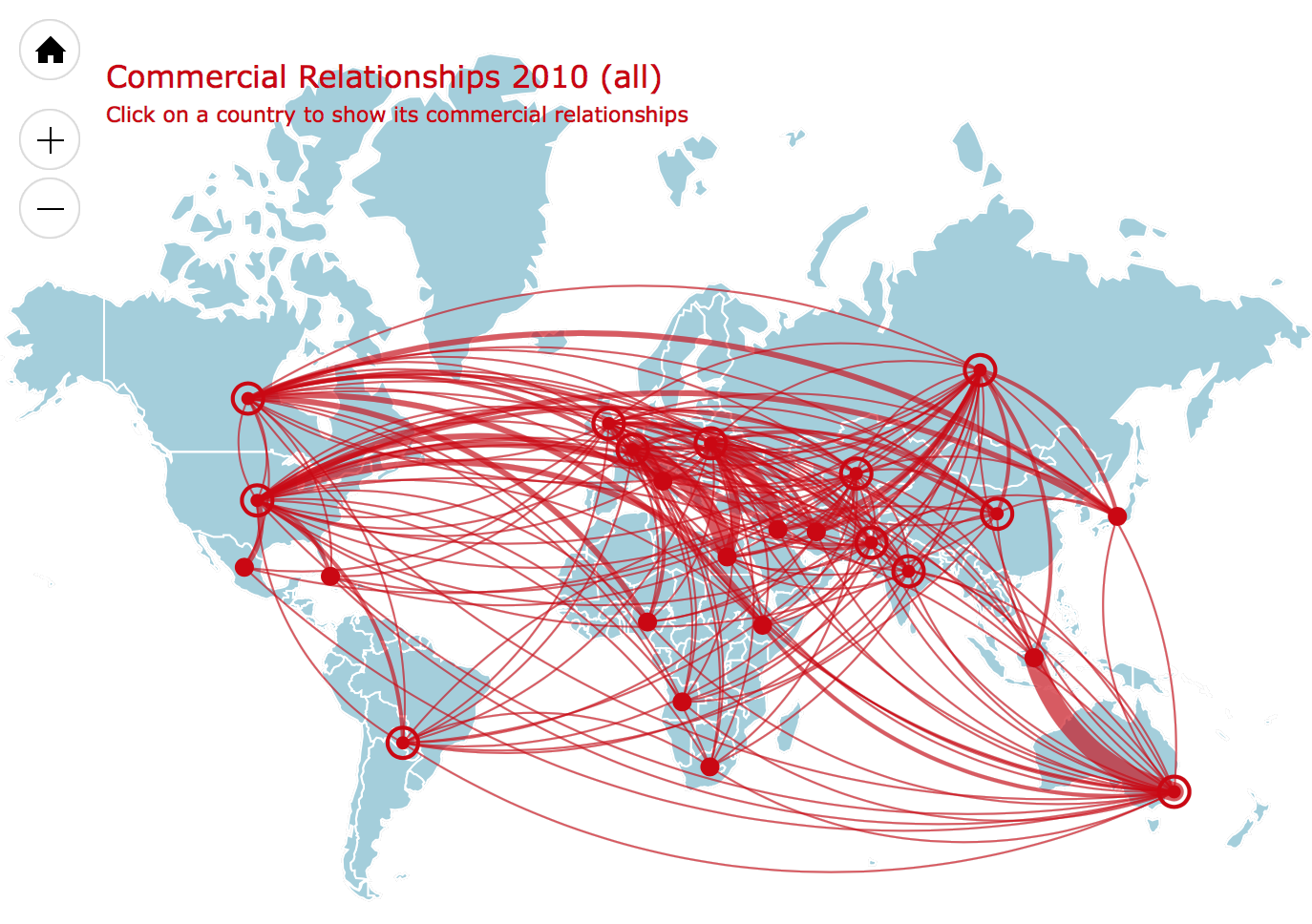}
	\vskip-3mm
	\raisebox{-0.7mm}{\includegraphics[scale=0.3]{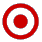}} {\tiny region with outgoing and incoming hub} \hskip3mm \raisebox{-0.0mm}{\includegraphics[scale=0.3]{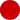}} {\tiny region with incoming hub only}
	\caption{Commercial relationships from simulation output for 2010.}
	\label{fig:map_2010}
\end{minipage}
\hskip1cm
\begin{minipage}[t]{7.5cm}
	\centering
	\includegraphics[width=10.5cm]{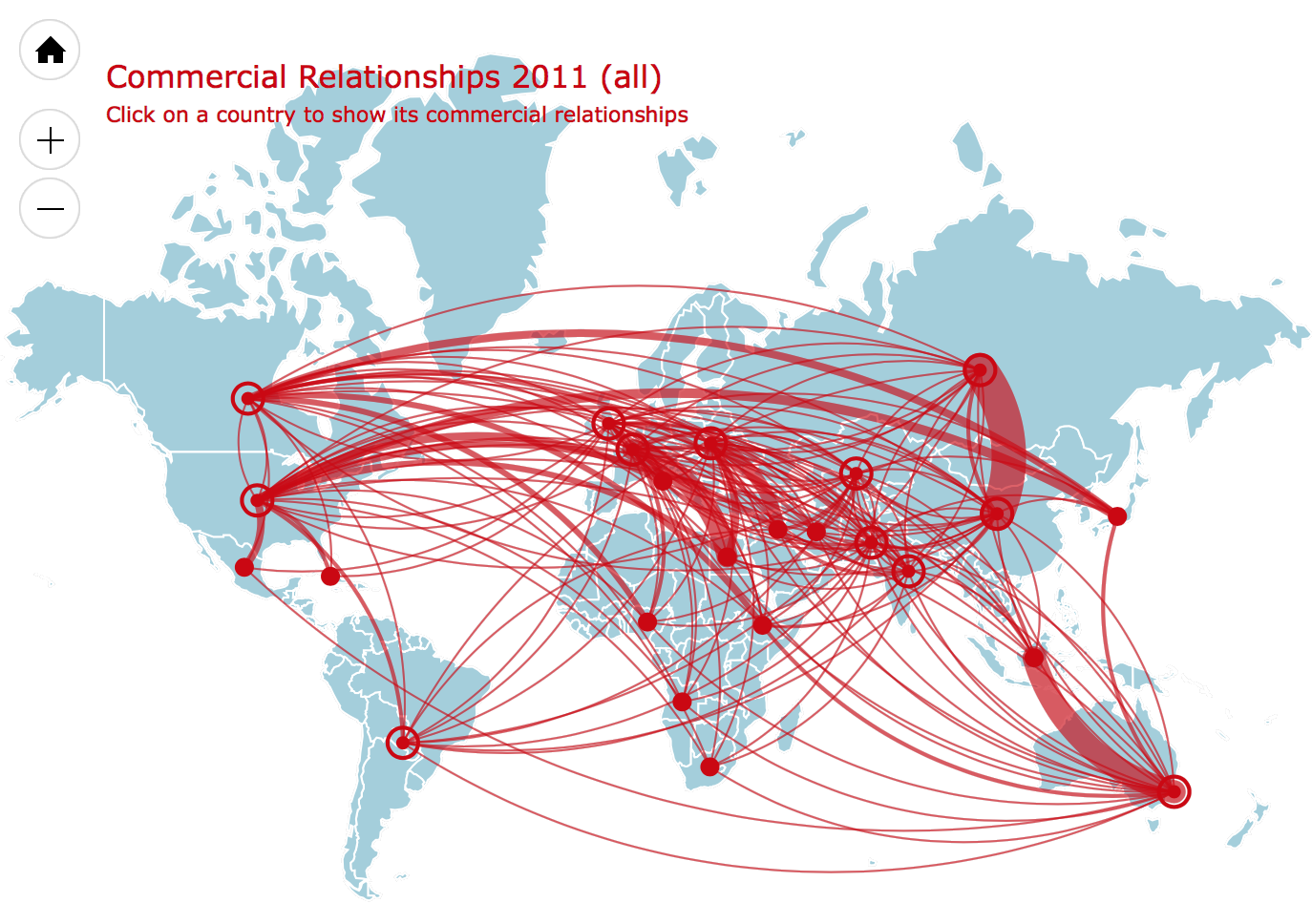}
	\vskip-3mm
	\raisebox{-0.7mm}{\includegraphics[scale=0.3]{buyer_producer_symbol.png}} {\tiny region with outgoing and incoming hub} \hskip3mm \raisebox{-0.0mm}{\includegraphics[scale=0.3]{buyer_symbol.png}} {\tiny region with incoming hub only}
	\caption{Commercial relationships from simulation output for 2011.}
	\label{fig:map_2011}
\end{minipage}
\end{figure}

On the quantity side, the model generates detailed data that allows for stimulating exercises, even though the scarcity of empirical data makes it difficult to validate the results. An example is the possibility to build the network of international exchanges and the possibility to trace the dynamics of such a network. This is a recent topic of investigation \citep{barigozzietal,fairetal}.   
We perform this exercise and report the result in Figure \ref{fig:map_2010}. The figure provides a visual representation of the international exchanges generated by the model for 2010. We recall that, in this type of representation, flows move clockwise. In other words a node A provides resources to a node B if moving from A along the edge to B implies a clockwise movement.
Some examples can help clarifying. According to Figure \ref{fig:map_2010}, South-Eastern Asia imported a relevant amount of wheat from Oceania (note the counter clockwise outer direction of the edges starting from Indonesia), while United States and Canada exported to several other regions (clockwise outer direction of many edges exiting from these countries).

A comparison of Figure \ref{fig:map_2010} and \ref{fig:map_2011} shows how the scenario can change year after year.
However, the evolution of international relationships generated by our model can be better appreciated visiting the web page \url{http://erre.unich.it/wheat_map/}. 

\subsection{Assessing the effects of the 2010 Russian Federation export blocking}

One of the most challenging use of the model presented in this paper is the evaluation of policy choice effects both on wheat prices and exchanged quantities. 
The Russian Federation export ban mentioned in the introduction (paragraph \ref{par:russia}) provides an occasion in this direction. 

In 2010 
a severe climate anomaly interested Eastern Europe causing many impacts related to heat-waves \citep{Barriopedro2010}, wildfires \citep{Liou2013} and air pollution episodes \citep{Konavalov}. In particular, Russian Federation, Kazakhstan, and Ukraine (all three amongst the world's top-10 wheat exporters) suffered the worst heatwave and drought in more than a century, while the Republic of Moldova was struck by floods and hail storms \citep{Arpe2011, Winne2016}. Furthermore, from early July to September a large crop production area was hit by wildfire with a significant production cut and only in Russian Federation grain yield was reduced by a third \citep{Liou2013}. As a consequence, in mid-August 2010 Russian Federation banned the export of domestically produced wheat \citep{Liou2013}. The measure was introduced to ensure wheat availability to domestic users after the dramatic crops loss. At the same time must be take into account that wheat prices have soared by about 90 percent since June 2010 because of these events as well as floods in Canada, while ban pushed price even higher.

As explained above, in our model import/export policies can be managed. This gives us a chance to observe how the system would have evolved if the export ban would have not be imposed. It is worth mentioning that because in the real world the ban indeed happened, it was active in our model during the calibration process. Using the calibrated parameters, the model was run two additional times: with the Russian Federation export ban enabled and disabled. 

\begin{figure}[ht]
	\centering
	\includegraphics[scale=1.0]{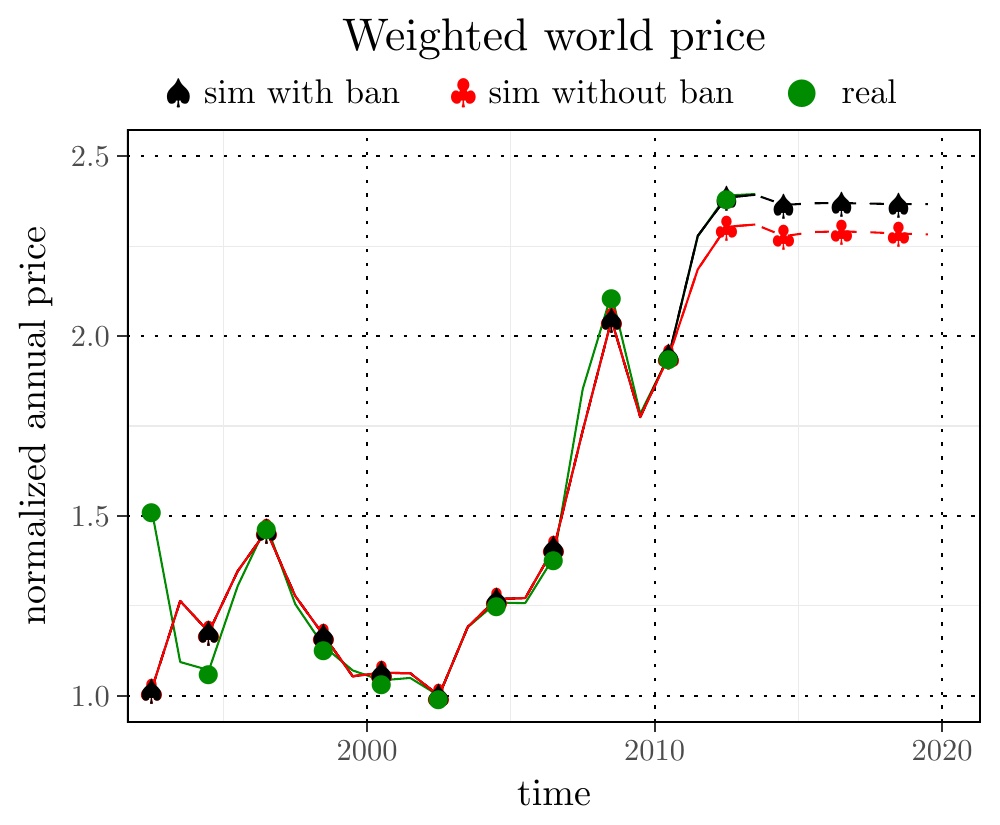}
	\caption{World weighted price with and without 2010 Russian Federation ban.}
	\label{fig:price_ban}
\end{figure}

Figure \ref{fig:price_ban} shows the result of this exercise on prices. According to the model output, wheat average world price would have been significantly lower in 2011-2013 if Russian Federation would not have imposed the ban (red line). Precisely, in 2013 the normalized observed and simulated (with ban) prices are basically the same and equal to 2.395. The normalized price in the simulation without ban is 2.31, with a -3.55\% deviation from the observed price.  
The figure also shows the projection of prices for the following years (dashed lines). They are obtained under the simplification that produced and demanded quantity keep constant to the 2013 levels. According to the model output the price gap tends to be constant in time. Although it is not easy to assess the robustness of this result due to the peculiar assumption under which it is obtained, the change in the level of prices observed in the dashed portion provides an opportunity for an additional observation. Since quantities are set exactly the same as in 2013, then the prices should not move from the 2013 level unless the system is not in its steady state. It follows that normally the system is not in its steady state and out of equilibrium situations often prevail.    

The effect of the ban on quantities can be evaluated by looking again at Figures \ref{fig:map_2010} and \ref{fig:map_2011}. According to the model, the export ban has modified Russian Federation commercial relationships. Figure \ref{fig:map_2011}, for example, makes it evident the relevant increase of the Russian Federation-China commercial relationship after the export ban was removed. Looking at the supporting material at \url{http://erre.unich.it/wheat_map}  it is possible to see how this link gradually normalizes in the following years.

\section{Conclusions and Further developments}\label{sec:conclusion}
In a globalized economy, the price of commodities, especially those of agricultural products depends on several factors: production technologies, commercial relationships, climate change forcing in various places of the globe only to cite the most important. 

Traditional analytic techniques find it difficult to take into account such variety of events. This work is a first step to take advantage of the agent-based techniques to handle all these factors. In particular, we aim at building a tool for analyzing the dynamics of cereals prices and exchanged quantities under alternative economic, environmental and climate conditions. 

In this paper, we have specialized the structure and the dynamics of an existing Agent based model for the generic analysis of commodities (the CMS model) to the wheat case. 
Changes are formulated in accordance with the economic determinants of Agricultural productions. Our variant of the original software is called the CMS-Wheat simulator.  

The careful model calibration we implement, together with the inclusion of crude oil price allow replication of empirical yearly weighted world price. By the quantity side, we have obtained promising results though further model building is needed. 

The model can be developed and improved in several directions.

\begin{figure}[!ht]
	\centering
	\includegraphics[scale=0.6]{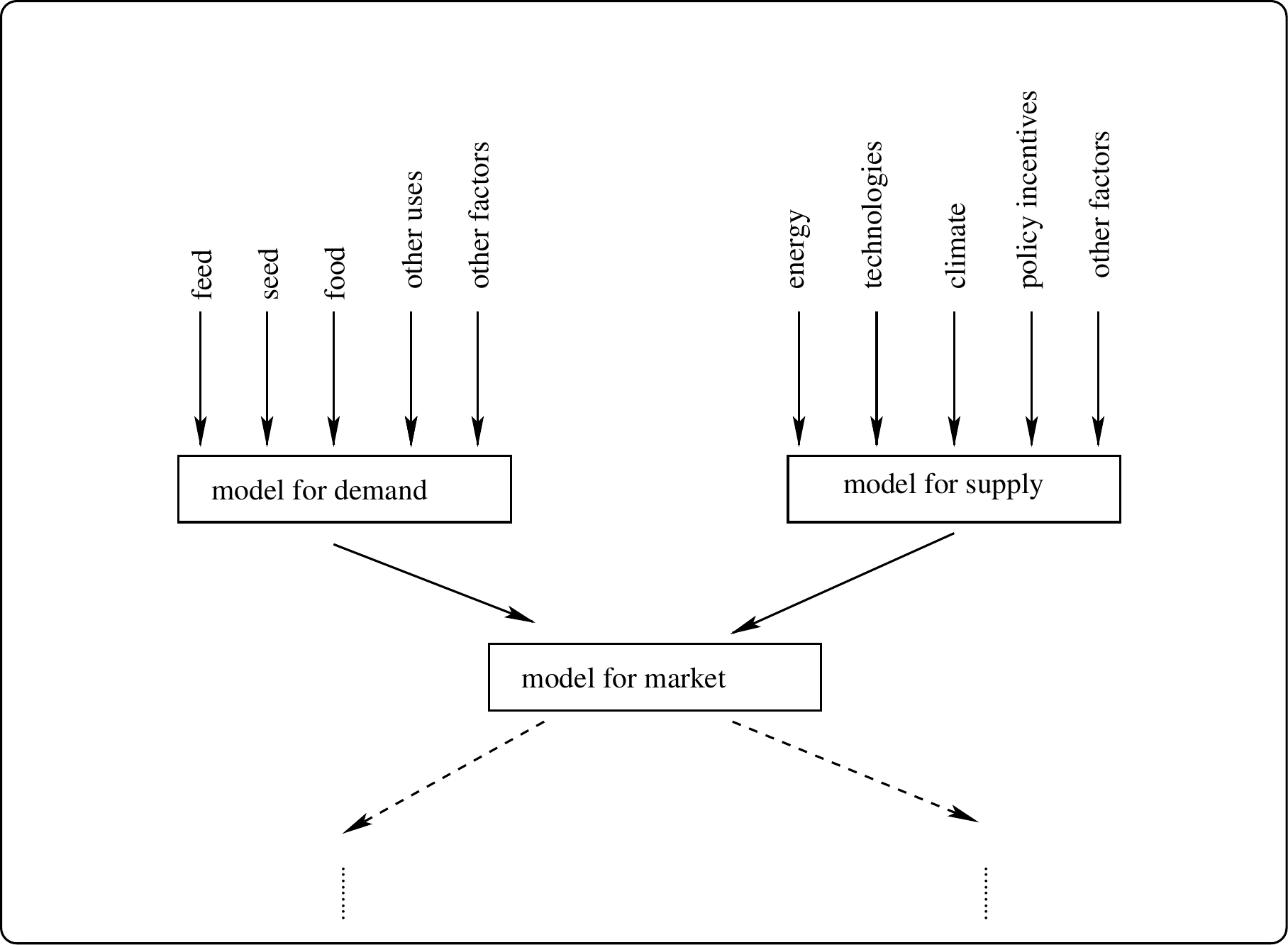}
	\caption{additional determinants of demand and supply to be considered}
	\label{fig:progress}
\end{figure}

Figure \ref{fig:progress} reports the top part of figure \ref{fig:model_visual} with some integration to highlight possible developments of the model.  
These developments involve either the demand side (differentiating by the utilization of cereals) and the supply side (conversion to biofuels, seasonal to decadal climate effects, technical levels, policy incentives, etc.). 

In particular, we think considering climate factors of primary importance in the further development of this work. 
We plan to develop the model to include climate variability forcing that is acting on wheat yields as  a primary exogenous factor. The combination of global scale climatic forcing, e.g. the  El Nin\~o-Southern Oscillation (ENSO), and local scale climatic characteristic, such as a period of drought, could modulate price dynamics or even  produce  shocks in the global market with relevant impacts. 
\cite{Iizumi2014-yh} and \citet{Gutierrez2017} find that the large scale atmospheric dynamics affect the local crop yields.  
Following this insight, we are presently running linear regression in order to identify the effect of the  ENSO on wheat yield of the top 20 worldwide producers. Preliminary results are encouraging: on average, we find 6-7 significant effects for each considered country (for a total of 138 significant relationships). All this will improve the modeling of wheat supply. 


Another important factor will be taken into account is the stock management of cereals that directly affects cereals international price. Traditionally stock-holding has been a private as well as a public activity. Private operations in this field are linked to the possibility of speculation based on future price expectations. On the other hand, Government agencies usually adopt a price band to balance supply and demand and to contain price volatility.
The management of the stock is strictly connected to the availability, access, utilization, and stability of food. 
It could represent the policy tool to reduce malnutrition in poorest countries.

All these developments will improve the performance of the model. It could therefore be used as a tool for producing reliable forecasts of prices and quantities both at global and region/country level. Another important goal of the model is to give recommendations about prevention policies that most reduce the negative effects of extreme negative events, emergency measures that implies less sacrifice for the population and measures to reduce the price volatility.



\section{Acknowledgements}
This study was partially developed in the framework of VOPA Project (VOlatilit\`a dei mercati e Produzioni Agricole: interazioni fra variabilit\`a climatica e sviluppo tecnologico delle nazioni nei meccanismi di formazione dei prezzi dei prodotti alimentari) and MACSUR (Modeling European Agriculture with Climate Change for Food Security - www.macsur.eu) - phase 2 Project.  

%
%

\bibliographystyle{chicago}
\bibliography{references}

\appendix
\section{Replicability and code availability}

The model used in this paper is available at the following link:\\ \url{https://github.com/gfgprojects/cms_wheat}\\ Both the source code and the documentation are available to interested people who wish to replicate the results or to develop the model.
Furthermore, the \verb+scripts+ folder provides all the R scripts \citep{R2018} used to generate the inputs for the model and for analyzing the simulations outcomes.  

The FAOSTAT dataset provides three levels of aggregation: country, regions and special groups.
Our data preparation process starts from the regions level. 
Regions are aggregation of countries. The regions level has itself three levels of aggregation.
The first one concerns partition of continents which include several countries. We will refer to this as the sub-continental level. The second one reports data from continents and the third one has data at world level. 
We start preparing the data using sub-continental data from Europe, Asia, Africa and Americas and we took directly continental data from Oceania.
Because some important producer countries deserve an individual treatment we further partitioned sub-continents to explicitly account for them. As shown in Table \ref{tab:zoneslist}, Northern America was broken down in the United States and its complement; Southern Asia in India, Pakistan and its complement; Eastern Asia in China and its complement; Eastern Europe in Russian Federation and its complement. This process lead us to partitioning the world in 24 geographic areas.   
\label{par:Aregions}
Each component of the wheat balance was recorded in a text file (.csv file extension). These files have 24 lines each of them relating to one geographic area. They are stored in the \verb+scripts/generate_inputs/data+ folder.
These data are further manipulated to obtain the files that are supplied as inputs to the ABM. The \verb+scripts/generate_inputs+ folder includes the \verb+R+ scripts which perform such manipulation and write files to the \verb+data+ folder where the ABM read inputs.

The motivation for this manipulation is that data recorded by FAO are the outcome of markets bargaining. Our model simulates markets functioning and delivers the outcome of market bargaining. Therefore, FAO and simulation results can be compared. Hence, FAO data cannot be used directly as inputs for our model, but they can be used to get back to wheat demand and supply which are the required inputs for the model.

A strategy would be to build a model for each of the balance component, i.e. a crop model for production, a selling strategy for the stock change and so on. However, this is a very demanding task that presents various difficulties.  Consider for example the food component of demand. It is straightforward to think that its main determinant is population, however real data show how the dynamics of population and the food component moves in totally different manner in several countries. While building a model for each component of FAO balances is a valuable strategy in the long run, in this paper we adopt a method allowing us to obtain demand components in a relatively easy way.   



\begin{figure}[h]
	\centering
\includegraphics[scale=0.4]{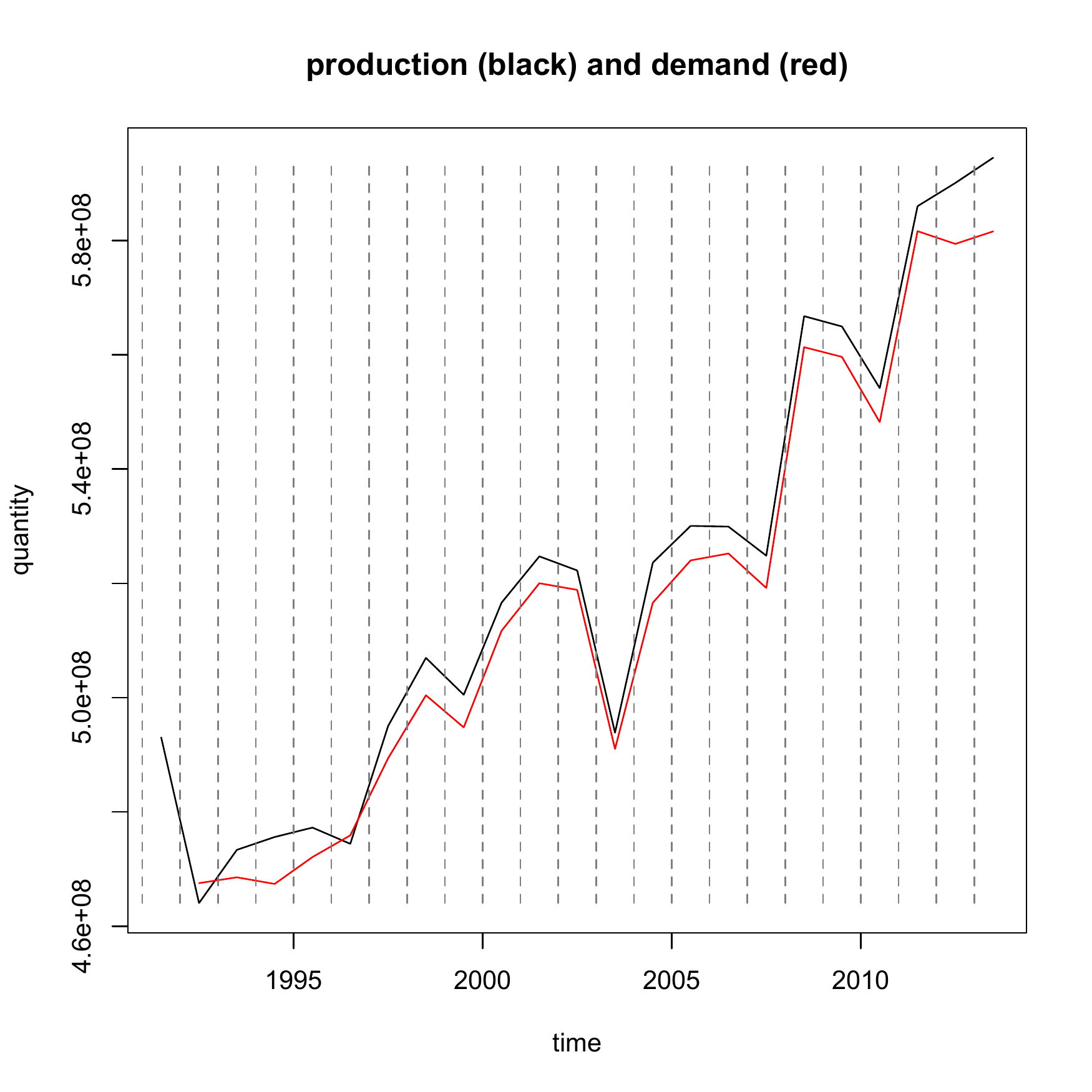}
	\caption{Unbalances in the data of FAOSTAT database.}
	\label{fig:world_unbalance}
\end{figure}

The idea behind the method was triggered by a check on FAOSTAT balances. If a sum over the balances of all the zones is performed, supplied and used quantities must coincide at global (world) level.
This check revealed a systematic deviation of supplied and used quantities as shown in the left chart of Figure \ref{fig:world_unbalance}. \label{par:world_unbalance}
This deviation is mainly due to accounting problems concerning internationally traded goods \citep{malovic13}. In fact, at the global level, total imports and total export should coincide. This equality is not respected in the data. In the following we will introduce a formal notation where we will use capital letters to denote yearly quantities supplied by the dataset. The global net import in year $t$ ($GNI_t$) defined by:
\[
	GNI_t=\sum_z (IMP_{z,t}- EXP_{z,t})
\]
where $z$ runs on all the considered regions, is not zero.
\newline
More precisely, if we compute the world supply in a given year $t$ as $Y_t=\sum_z Y_{z,t}$ and the world demand as $D_t=\sum_z D_{z,t}$ we have
\[
Y_t-D_t=-GNI_t
\]
The unbalance at the world level is shown figure \ref{fig:world_unbalance}.
\newline
We can now mine the data in order to ensure that world supply equals world demand, i.e $GNI=0$ in each period.
Starting from
\[
Y_t-D_t+GNI_t=0
\]
we can compute the mined level of supply, say $\hat{y}$, as the level which meets FAO demand
\[
	\hat{Y}_t-D_t=0
\]
Therefore
\[\hat{Y}_t=Y_t+GNI_t=\left(1+\frac{GNI_t}{Y_t}\right)Y_t=\left(1+\hat{\eta}^y_t\right)Y_t\]
Where $\hat{\eta}^y_t$ is a percentage deviation. Going back to figure \ref{fig:world_unbalance}, $\hat{Y}$ now overlaps the demand line.
\newline
On the other hand, we can compute the mined level of supply, say $\hat{D}$, as the level which meets FAO supply
\[
	Y_t-\hat{D}_t=0
\]
therefore
\[\hat{D}_t=D_t-GNI_t=\left(1-\frac{GNI_t}{D_t}\right)D_t=\left(1+\hat{\eta}^d_t\right)D_t\]
Where $\hat{\eta}^d_t$ is a percentage deviation.  Going back to figure \ref{fig:world_unbalance}, $\hat{D}$ now overlaps the supply curve.

The discussion in the previous paragraph shows that we can modify data in order to achieve a given result. In this case the goal was to obtain $GNI_t=0$.
This also provides a method to go back from FAO balance data to the unknown level of demand and supply. 
In this case, we choose the goal of replicating the yearly weighted world price. We therefore undertook the endeavor of finding supply and demand percentage deviations which would make the model to better achieve the goal. We will denote these deviations as $\tilde{\eta}^y_t$ and $\tilde{\eta}^d_t$.
We adopt the simplification $\tilde{\eta}^y_t=0 \ \forall t$ because supply is strongly influenced by production. We therefore focused on the  $\tilde{\eta}^d_t$s.

Once the $\tilde{\eta}^d_t$s has been set, we compute the desired demand for each country by computing the following value
\[\tilde{D}_{z,t}=(1-\tilde{\eta}^d_t)D_{z,t}\]

Based on the comparison between $\tilde{D}_{z,t}$ and $Y_{z,t}$ we partition regions in two sets: the net international buyers set and its complement. A region belongs to the net international buyers set if $\tilde{D}_{z,t}\ge Y_{z,t} \ \forall t$. The idea behind this classification is that all the production of a net international buyer is employed in domestic uses. Therefore, a net international buyer has not a production to be sold to other countries, hence it does not has an international market. 

Summing up, regions are partitioned in the set of net international buyers and the set of international supplier. International suppliers organize international markets where they offer their production $Y_{z,t}$. On the other hand, international suppliers direct their demand ($\tilde{D}_{z,t}$) to their own or other international markets. Net international buyers direct their net demand $\tilde{D}_{z,t}-Y_{z,t}$ to international markets.

This process allows us to reduce the number of international markets and to sift the market maker. \\ 
The \verb+r_reduce_number_of_producers_food.R+ script transforms the aggregated data reducing the number of producers.
The script outputs several files in the \verb+data+ folder. These files are loaded by the code in order to setup the simulation.   

It is worth noticing that the $\tilde{D}_{z,t}$s are given as input to the model and can be viewed as a country's desired quantities at an average price. The yearly data are then transformed in monthly $\tilde{d}_{z,t}$ values by the software at initialization time. The observed exchanges are in general different from those because a buyer will bought a higher (lower) quantity if the price is low (high).

\end{document}